\documentclass[prb,twocolumn,aps,showpacs,superscriptaddress,epsfig]{revtex4-2}
\usepackage[ruled,vlined]{algorithm2e}
\usepackage{tabularx}
\usepackage{array}
\usepackage{amssymb}
\usepackage{amsbsy}
\usepackage{bbm}
\usepackage{amsmath}
\usepackage{amsfonts}
\usepackage{bm}
\usepackage{physics}
\usepackage[dvipsnames,table]{xcolor}
\usepackage{boldline}
\usepackage{graphicx}
\usepackage[table]{xcolor}
\definecolor{Gray}{gray}{0.75}
\usepackage{epsfig}
\usepackage[bb=boondox]{mathalfa}
\usepackage{lipsum}
\usepackage{graphicx}
\usepackage{subfigure}
\usepackage[normalem]{ulem}
\usepackage{dsfont}

\usepackage[toc,page]{appendix}
\usepackage{comment} 
\usepackage{tikz}
\usetikzlibrary{
    arrows.meta,
    positioning,
    shapes.geometric,
    fit,
    calc
}
\usepackage{array}
\usepackage{mathrsfs}
\usepackage{soul}
\usepackage{float}
\usepackage{hyperref}
\hypersetup{
	colorlinks=true,
	linkcolor=blue,
	filecolor=cyan,      
	urlcolor=magenta,
	citecolor=red,
}

\newcommand{\nodag}{{\phantom{\dagger}}}

\makeatother
\makeatletter
\newcommand{\supplementarytableofcontents}{%
  \section*{Supplementary Contents}
  \@starttoc{stoc}%
}
\makeatother

\makeatletter
\newcommand{\suppTOCon}{%
  \let\old@addcontentsline\addcontentsline
  \renewcommand{\addcontentsline}[3]{%
    \edef\@tempa{##1}\edef\@tempb{toc}%
    \ifx\@tempa\@tempb
      \old@addcontentsline{stoc}{##2}{##3}%
    \else
      \old@addcontentsline{##1}{##2}{##3}%
    \fi
  }%
}
\newcommand{\suppTOCoff}{%
  \let\addcontentsline\old@addcontentsline
}
\makeatother

\newcommand{\supplement}[1]{%
  \clearpage%
  \onecolumngrid%
  \title{#1}%
  \maketitle%
  \setcounter{section}{0}%
  \setcounter{subsection}{0}%
  \setcounter{equation}{0}%
  \setcounter{figure}{0}%
  \setcounter{table}{0}%
  \setcounter{page}{1}%
  \makeatletter%
  \renewcommand{\thesection}{S\arabic{section}}%
  \renewcommand{\thesubsection}{\Alph{subsection}}%
  \renewcommand{\thepage}{S\arabic{page}}%
  \numberwithin{figure}{section}%
  \numberwithin{table}{section}%
  \numberwithin{equation}{section}%
  \renewcommand{\thefigure}{\thesection.\arabic{figure}}%
  \renewcommand{\thetable}{\thesection.\arabic{table}}%
  \renewcommand{\theequation}{\thesection.\arabic{equation}}%
  \makeatother%
}

\makeatletter
\def\maketitle{%
  \@author@finish%
  \title@column\titleblock@produce%
  \suppressfloats[t]%
}
\makeatother

\usepackage{makerobust}
\newcommand{\makeauthor}[2]{\newcommand{#1}[1]{{%
  \protect%
  \color{#2}{%
    \bfseries%
    \begingroup\escapechar=-1\edef\x{\endgroup\string#1}\x:%
  }\itshape{} ##1}}%
  \MakeRobustCommand#1}
\makeauthor{\mk}{Plum}
\makeauthor{\sm}{ForestGreen}
\makeauthor{\js}{orange}
\makeauthor{\ji}{blue}
\makeauthor{\dr}{cyan}
\makeauthor{\rt}{green}
\makeauthor{\pw}{purple}
\makeauthor{\todo}{red}

\begin{document}

\title{Slave-boson Formalism for Superconducting Pairing at Strong Coupling}

\author{Sarbajit Mazumdar}
\affiliation{Institut f\"ur Theoretische Physik und Astrophysik and W\"urzburg-Dresden Cluster of Excellence ctd.qmat, Julius-Maximilians-Universit\"at W\"urzburg, Am Hubland, Campus S\"ud, W\"urzburg 97074, Germany}
\author{Jonas Issing}
\affiliation{Institut f\"ur Theoretische Physik und Astrophysik and W\"urzburg-Dresden Cluster of Excellence ctd.qmat, Julius-Maximilians-Universit\"at W\"urzburg, Am Hubland, Campus S\"ud, W\"urzburg 97074, Germany}
\author{Jannis Seufert}
\affiliation{Institut f\"ur Theoretische Physik und Astrophysik and W\"urzburg-Dresden Cluster of Excellence ctd.qmat, Julius-Maximilians-Universit\"at W\"urzburg, Am Hubland, Campus S\"ud, W\"urzburg 97074, Germany}
\author{David Riegler}
\affiliation{Institut f\"ur Theorie der Kondensierten Materie, Karlsruhe Institute of Technology, D-76128 Karlsruhe, Germany}
\affiliation{Institut f\"ur Theoretische Physik und Astrophysik and W\"urzburg-Dresden Cluster of Excellence ctd.qmat, Julius-Maximilians-Universit\"at W\"urzburg, Am Hubland, Campus S\"ud, W\"urzburg 97074, Germany}
\author{Peter W\"olfle}
\affiliation{Institut f\"ur Theorie der Kondensierten Materie, Karlsruhe Institute of Technology, D-76128 Karlsruhe, Germany}
\author{Ronny Thomale}
\affiliation{Institut f\"ur Theoretische Physik und Astrophysik and W\"urzburg-Dresden Cluster of Excellence ctd.qmat, Julius-Maximilians-Universit\"at W\"urzburg, Am Hubland, Campus S\"ud, W\"urzburg 97074, Germany}
\author{Michael Klett}
\email{michael.klett@uni-wuerzburg.de}
\affiliation{Institut f\"ur Theoretische Physik und Astrophysik and W\"urzburg-Dresden Cluster of Excellence ctd.qmat, Julius-Maximilians-Universit\"at W\"urzburg, Am Hubland, Campus S\"ud, W\"urzburg 97074, Germany}

\date{\today}

\begin{abstract}
We study the emergence of superconductivity in the one-band Hubbard model using the spin-rotation-invariant Kotliar–Ruckenstein slave-boson (SB) approach. Motivated by its intrinsically renormalized mean-field ground state, we construct an effective pairing vertex from dynamical fluctuations about the saddle point. Solving the anisotropic, frequency-dependent gap equation on the square lattice, we map the pairing instabilities across doping, interaction, temperature and real-frequency gap structure that qualitatively match experimental cuprate observations. This framework merges strong-correlation SB-type renormalizations with RPA-type pairing transparency, providing a scalable route to modeling multi-orbital superconductivity at strong coupling.
\end{abstract}

\maketitle

\section{Introduction}



Strongly correlated electron systems host a broad range of emergent phases, including charge and spin ordered states, unconventional superconductivity, and correlated metals with properties beyond conventional order. Strategies for solving this analytically non-tractable problem center around (i) model simplification and (ii) designing numerical ways to address the strong electronic interactions. A central minimal setting is the single-band Hubbard model on the square lattice with nearest- and next-nearest-neighbor hopping, widely used as a controlled arena for cuprate-inspired physics. Across complementary numerical approaches, robust evidence has accumulated for dominant $d$-wave pairing tendencies near intermediate fillings, including quantum Monte Carlo (QMC)\cite{Blankenbecler1981m,Zhang1997p,Cao2025d}, density-matrix renormalization group (DMRG) studies on ladders and wide cylinders \cite{White1992d,White2003s,Jiang2024g}, density-matrix embedding theory (DMET) \cite{Knizia2012d,Zheng2016g}, and renormalization-group (RG) benchmarks \cite{Shankar1994r,Halboth2000d,Honerkamp2001m,Metzner2012f}. At the same time, a unified description that connects pairing symmetry, fluctuation spectra, and dynamical gap structure across the broad interaction and doping ranges of interest remains an open challenge.

A key conceptual milestone is the observation that superconductivity can arise even from purely repulsive interactions by Kohn and Luttinger \cite{Kohn1965n,Luttinger1966n}. In lattice settings, the prevalent physical picture is that pairing is promoted by the exchange of collective fluctuations, most prominently spin fluctuations. Within this framework, Berk and Schrieffer, along with later works by Scalapino et al., established the connection between enhanced magnetic response and unconventional pairing \cite{Berk1966e,Scalapino1986d,Scalapino1995t}. Random-phase-approximation (RPA)-based treatments make this mechanism particularly transparent yet constrain their ability to reproduce many qualitative features of the ensuing phase diagrams to weak-to-intermediate-coupling \cite{Romer2015p,Simkovic2016g,Monthoux1994se,Bickers1989c}. In particular for intermediate and strong coupling, the predictive power of such approaches hinges on the accuracy of the underlying dynamical susceptibilities, as capturing both their strong-correlation renormalizations and their full frequency dependence is essential for controlled statements about pairing kernels and the resulting frequency-dependent order parameter.

Gauged by field activity, dynamical mean field theory (DMFT) \cite{Metzner1989c,Georges1992h,Georges1996d,Held2007e} lends itself as a preferred modern-day numerical approach to address such a problem outset of correlated electron systems, where strong dynamical fluctuations are crucial and assumed to be mainly local. As unconventional pairing specifically entails non-local correlators emanating from non-local fluctuations, however, it is desirable to achieve a methodological ansatz which generically incorporates dynamical non-local fluctuations and is numerically efficient to allow large scale scans to access multi-parametric phase diagrams. While significant progress has been made within DMFT to include non-local correlations \cite{Rohringer2018d}, it does remain rather involved in terms of technicality and numerical effort, and hence keeps challenging the quantitative access to non-local phenomena of strongly correlated electron systems such as unconventional superconductivity.

The Gutzwiller wave function \cite{Gutzwiller1963e,Gutzwiller1964e,Gutzwiller1965c} offers a compact variational route to incorporate correlations while retaining itinerancy. Its key insight is that by suppressing costly double occupancies via a local projector, it reduces the probabilistic weight of hopping processes that would create them, thereby inherently capturing bandwidth renormalization and Mott physics \cite{Brinkman1970a,Vollhardt1984n}. 
The Kotliar–Ruckenstein slave-boson (KRSB) formulation translates this variational insight into an explicit fermionic operator framework by introducing auxiliary bosons that resolve the local occupancy configurations (empty, singly, doubly occupied), recovering the Gutzwiller approximation at the mean-field saddle point \cite{Kotliar1986n}. Its spin-rotation-invariant generalization \cite{Fresard1997i,Lechermann2007r} enables a consistent treatment of commensurate and incommensurate collective spin and charge dynamics as well as renormalizing the underlying band structure \cite{Legner2014t,Klett2020t}. Importantly, Gaussian fluctuations around this correlated saddle point yield fully dynamical susceptibilities that already encode strong-correlation renormalizations at both quasiparticle and vertex levels—originating from a correlated reference state rather than bare-band propagators. This marks a conceptual departure from RPA-type treatments, providing a starting point that already incorporates medium to strong-correlation at the quasiparticle and vertex level for the subsequent pairing analysis.  The underlying SRIKR framework has furthermore been benchmarked
extensively against complementary many-body methods. Comparisons of the
ground-state energy, double occupancy, magnetization, and spin and charge
structure factors with auxiliary-field quantum Monte Carlo, DMET, variational calculations, and related numerical
approaches show good agreement over broad parameter regimes
\cite{Riegler2020s,Zimmermann1997s,Riegler2026f}. These benchmarks provide
independent support for the correlated quasiparticle background and
collective response functions used as the starting point of the present
pairing construction.


In this Article, we leverage dynamical slave-boson susceptibilities to construct an effective two-particle pairing vertex that is explicitly antisymmetric under fermionic exchange and decomposes into density/spin and singlet/triplet channels. Using this vertex, we solve the anisotropic, frequency-dependent gap equation on the square lattice and classify leading pairing symmetries across doping, interaction strength, and temperature, revealing spin-fluctuation dominance and $d$-wave phase evolution in quantitative agreement with available numerical benchmarks and experimental observations in cuprates. We find that the pairing kernel is generically dominated by the spin-fluctuation channel, yielding a systematic superconducting symmetry map, and extract real-frequency gap features via Pad\'e continuation of Matsubara-axis solutions. The framework we unfold here provides a direct, scalable route from slave-boson fluctuation dynamics to superconducting pairing, suggests immediate applicability to strong coupling, and renders itself naturally extensible to multi-orbital settings.

The article is organized as follows. In Section \ref{sec:mm} we introduce the principal methodological and model starting point, as we particularize to the single band Hubbard model on the square lattice and expand on the elementary properties of the effective two-particle vertex and the superconducting gap equation from the viewpoint of slave boson mean field theory. For the sake of readability, most technical details are provided in the
Supplemental Material~\cite{SupplementalMaterial}. Section \ref{sec:re} continues by elaborating on the results we have achieved, starting with the spin/charge fluctuation profile, the phase diagram as a function of filling $n$ and Hubbard $U$. This is then followed up by moving to superconductivity, where we resolve the filling dependence of superconductivity seeded by strong coupling fluctuations, the critical temperature, and the dynamical gap profile. We conclude in Section \ref{sec:co} that our framework establishes a promising new angle at strong coupling superconductivity, and outline several research threads for future work. 

\section{Model \& Method}\label{sec:mm}
We study the single-band Hubbard model on the two-dimensional square lattice, which captures the fundamental competition between electron itinerancy and strong local correlations through nearest-neighbor hopping and on-site Coulomb repulsion \cite{Hubbard1963e,Kanamori1963e,Gutzwiller1963e}. This model serves as a paradigmatic framework for understanding emergent phenomena in correlated electron systems, providing a minimal yet nontrivial setting in which magnetism, Mott insulating behavior, and unconventional superconductivity naturally arise. In the context of cuprate superconductors, effective one-band descriptions of the CuO$_2$ planes have established the Hubbard model as a widely used minimal starting point \cite{Anderson1987t,Zhang1988e}. Around half filling, the model exhibits antiferromagnetic order, and upon doping, it develops strong magnetic fluctuations that can promote $d$-wave pairing. Its properties have been extensively benchmarked across complementary numerical methods, ensuring a reliable reference for further methodological development and comparison. The corresponding Hamiltonian is given by,
\begin{align}
\mathcal{H} &= -\sum_{ij\sigma} t_{ij}\, c^{\dagger}_{i\sigma} c_{j\sigma}
+ U \sum_i n_{i\uparrow} n_{i\downarrow}
- \mu \sum_{i\sigma} n_{i\sigma}
\nonumber \\
&= \sum_{\mathbf{k}\sigma} (\epsilon_{\mathbf{k}} - \mu)\, c^{\dagger}_{\mathbf{k}\sigma} c_{\mathbf{k}\sigma}
+ U \sum_i n_{i\uparrow} n_{i\downarrow}.
\label{eq:Hubbard}
\end{align}
where $c^{\dagger}_{i\sigma}$ ($c_{i\sigma}$) creates (annihilates) an electron with spin $\sigma$ on lattice site $i$. Analogously, $c^{(\dagger)}_{\mathbf{k}\sigma}$ denotes the Fourier transform of $c^{(\dagger)}_{i\sigma}$ and acts in momentum space. Here, $\epsilon_{\mathbf{k}}$ is the single-particle dispersion defining the band structure, $\mu$ is the chemical potential controlling the electron filling, and $n_{i\sigma} = c^{\dagger}_{i\sigma} c_{i\sigma}$ is the number operator for electrons with spin $\sigma$ at site $i$. The nearest-neighbor hopping sets the energy unit, $t$, and we also include a next-nearest-neighbor hopping $t'$. All energy scales are given in units of $t$. The Gutzwiller projection, a well-established variational mean-field approach capturing Mott physics and bandwidth renormalization, serves as our starting point~\cite{Brinkman1970a,Vollhardt1984n}. We build upon it via the spin-rotation-invariant Kotliar--Ruckenstein (SRIKR) slave-boson representation, enabling Gaussian fluctuations around the paramagnetic saddle to yield physically motivated dynamical susceptibilities for the pairing vertex. The physical electron is 
\begin{equation}
c^\dagger_{i\sigma} = \sum_{\sigma'} z^\dagger_{i,\sigma\sigma'} f^\dagger_{i,\sigma'}, 
\end{equation}
with pseudofermions $f^{(\dagger)}_{i\sigma}$ and bosons $e_i,d_i,p_{i\tau}$ ($\tau=0,1,2,3$) projecting local occupancies (empty, double, single). The renormalization matrix $z(e,p,d)$ encodes Gutzwiller factors, preserves SU(2) spin rotation symmetry, and recovers the noninteracting limit. Local constraints restrict to the physical Hilbert space; we refer to the corresponding Lagrange multipliers and bosons representing occupancies as bosonic fields $\psi$. To obtain response functions, we consider Gaussian fluctuations about the paramagnetic saddle point~\cite{Li1991d,Zimmermann1997s} and expand the action ($\mathcal{S}$) to quadratic order in the bosonic fields $\psi_{\mu}$:

\begin{equation}
\delta \mathcal{S}^{(2)} \;=\; \sum_{\mathbf q,n} \delta \psi_{\mu}(-q)\,
\mathcal{M}_{\mu\nu}(q)\,
\delta \psi_{\nu}(q),
\label{eq:deltaS}
\end{equation}

with fluctuation matrix
\begin{equation}
\mathcal{M}_{\mu\nu}(q)
:= \frac{\delta^2 \mathcal{S}(\psi)}%
{\delta \psi_{\mu}(-q)\,
 \delta \psi_{\nu}(q)},
\label{eq:Mmatrix}
\end{equation}

where $q=(\mathbf{q}, i\Omega_n)$ and $\Omega_n=2\pi n T$ ($n\in\mathds{Z}$) is a bosonic Matsubara frequency. From this we can compute charge and spin susceptibilities. The spin (s) and charge (c) susceptibilities arising from the fluctuation expansion are linear combinations of different elements of the inverse fluctuation matrix $\mathcal{M}$
\begin{equation}
\chi_{s/c}(q) =\sum_{\mu,\nu} \mathscr{C}^{\mu,\nu}_{s/c} \mathcal{M}^{-1}_{\mu\nu}(q). 
\label{eq:11a}
\end{equation}
where $\mathscr{C}^{\mu,\nu}_{s/c}$ are susceptibility-dependent form factors. The static spin and charge susceptibilities are obtained at the lowest bosonic Matsubara frequency, \(i\Omega_0 = 0\):
\begin{equation}
\chi_{s/c}(\mathbf{q}) = \chi_{s/c}(\mathbf{q}, i\Omega_0).
\end{equation}
We further introduce the bare susceptibility $\chi_{0}(q)$ as the Lindhard bubble of renormalized pseudofermion bands. Details can be found in the Supplemental
Material~\cite{SupplementalMaterial}, Sec.~\ref{sec:sb}, as well as in
Refs.~\cite{Riegler2020s,Seufert2021b}.

\subsection{Pair interaction}
\label{vertex}

To access the Cooper channel generated by a purely repulsive local interaction,
we integrate out particle-hole fluctuations and work with the resulting
two-particle irreducible pairing kernel. In the spirit of the Kohn--Luttinger
construction \cite{Kohn1965n,Luttinger1966n}, the leading nontrivial contribution arises at
second order in interaction from the exchange of collective spin and
charge modes. In the present approach these modes are encoded by the SB
susceptibilities $\chi_s$ and $\chi_c$, which therefore provide the natural
building blocks for the fluctuation-mediated effective pairing vertex (following a RPA-like resummation \cite{Berk1966e,Scalapino1986d,Scalapino1995t}) used below. In order to construct the effective pair interaction derived from the exchange of spin or charge fluctuations, we need to know the vertex functions $\mathcal{K}_{s,c}(q)$ coupling the quasiparticles to the fluctuations. In diagrammatic language, these are given by the particle-hole irreducible vertex functions in the respective channels \cite{Pfitzner1985}. Within the slave-boson framework, we
identify their effective counterparts by rewriting the susceptibilities in an
RPA-like form,
\begin{equation}
\chi_{s,c}(q)
=
\frac{\chi_0(q)}
{1-\mathcal{K}_{s,c}(q)\chi_0(q)} ,
\label{eq:chi_K_relation}
\end{equation}
where $\chi_0(q)$ is the Lindhard susceptibility of the renormalized
slave-boson mean-field quasiparticles. Inverting Eq.~\eqref{eq:chi_K_relation}
gives
\begin{equation}
\mathcal{K}_{s,c}(q)
=
\frac{\chi_{s,c}(q)-\chi_0(q)}
{\chi_{s,c}(q)\chi_0(q)} .
\label{eq:Ksc_def}
\end{equation}
This procedure extracts the effective quasiparticle--fluctuation
vertices \(\mathcal K_{s,c}\) from the slave-boson response functions.
The Cooper-channel pairing interaction is then constructed from the
exchange of the corresponding collective propagators. With the normalization used here, the singlet and triplet pairing $\mathcal{V}_{s/t}(q)$ kernels can
then be written as
\begin{align}
\mathcal{V}_s(q)
&=
\mathcal{K}_{\mathrm{irr}}(q)
+\frac{3}{2}\mathcal{K}_s^2(q)\chi_s(q)
-\frac{1}{2}\mathcal{K}_c^2(q)\chi_c(q).
\label{eq:Vs_main}
\\
\mathcal{V}_t(q)
&=
-\frac{1}{2}\mathcal{K}_s^2(q)\chi_s(q)
-\frac{1}{2}\mathcal{K}_c^2(q)\chi_c(q),
\label{eq:Vt_main}
\end{align}
See Supplemental Material~\cite{SupplementalMaterial},
Sec.~\ref{sec:supp_vertex_decoupling}, for more details. Here $\mathcal{K}_{\mathrm{irr}}(q)$ denotes the residual fully irreducible contribution
to the singlet Cooper-channel interaction, i.e., the part not generated by
exchange of spin or charge fluctuations. In the bare Hubbard RPA limit this
term reduces to the first-order onsite repulsion, $\mathcal{K}_{\mathrm{irr}}=U$. In the
present slave-boson formulation it is renormalized by the correlated
quasiparticle background. Since our focus is on unconventional sign-changing
pairing channels, this local residual contribution does not
drive nor hinder such instabilities. We can therefore just absorb its scale
into the effective onsite interaction when analyzing the
fluctuation-mediated part of the pairing kernel. In the later
evaluations we will focus on the exchange of spin fluctuations as the principal mechanism of superconductivity. We assess an effective onsite scale by evaluating the spin vertex at the
dominant magnetic wave vector, $
U_{\mathrm{eff}}
=
\mathcal{K}_s(\mathbf Q,0)$, where $\mathbf Q$ denotes  the wave vector (including symmetry-related equivalents) at which the static spin response
$\chi_s(\mathbf q,0)$ is maximal. Equivalently,
\begin{equation}
U_{\mathrm{eff}}
=
\frac{\chi_s(\mathbf Q,0)-\chi_0(\mathbf Q,0)}
{\chi_s(\mathbf Q,0)\chi_0(\mathbf Q,0)} .
\label{eq:Ueff_chi}
\end{equation}
This definition ties the pairing scale to the strongest magnetic fluctuation
channel of the slave-boson paramagnetic reference state. The explicit singlet/triplet projection of the antisymmetrized two-particle
vertex contains both transferred four-momenta $k-k'$ and $k+k'$ (see Supplemental Material~\cite{SupplementalMaterial},
Sec.~\ref{sec:supp_vertex_decoupling}). In the present
one-band, inversion-symmetric case with zero center-of-mass Cooper pairing and
gap functions of definite singlet/triplet parity, the $k+k'$ contribution maps
onto the $k-k'$ contribution inside the full Brillouin-zone gap equation \cite{Scalapino1995t}.

\begin{figure*}
    \centering 
     \includegraphics[width=1\textwidth]{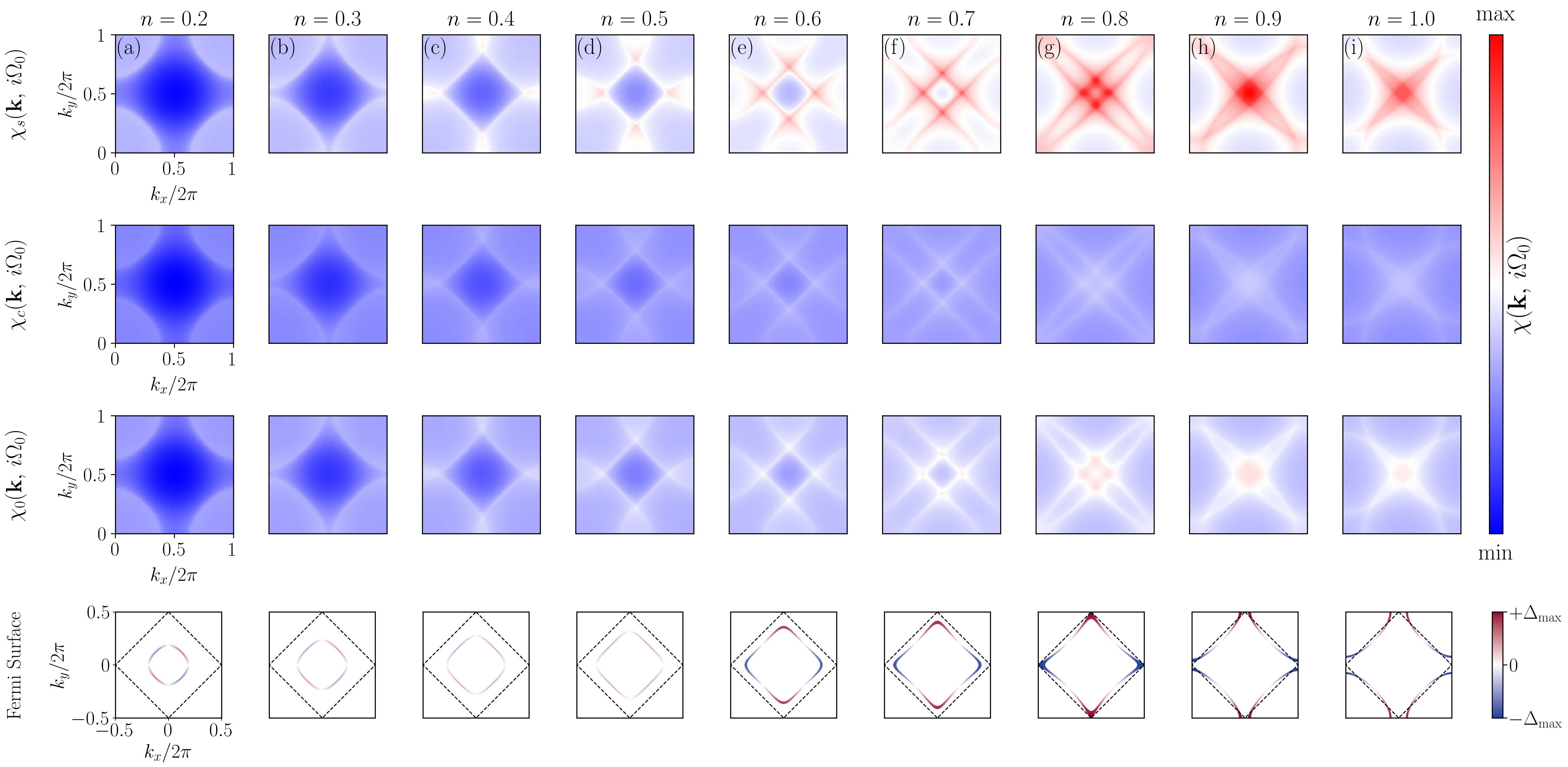}
     \vspace{-1\baselineskip}
\caption{
From top to bottom: static spin, charge, and bare susceptibilities, $\chi_{a}(\mathbf{k}, i\Omega_{0})$ with $a=s,c,0$, and the leading singlet superconducting instability projected onto the Fermi surface. The panels correspond to fillings $n=0.2$ to $1.0$ [(a)–(i)]. The susceptibility plots are shown on the same color scale to allow direct comparison. In all cases, the spin response exceeds the charge response, indicating spin-fluctuation-dominated pairing. The antiferromagnetic (AF) zone boundary is marked by a black dashed line.
}

    \label{fig:1}
\end{figure*}

\subsection{Static \& Dynamic Gap equations}

At finite temperature, the superconducting gap can be obtained from a
frequency-dependent (Eliashberg-type) gap equation \cite{Eliashberg1960i,Monthoux1994se,Bickers1989c}. For an effective interaction
in the singlet/triplet channel, $\mathcal{V}_{s/t}(\mathbf{p}, i\Omega_{nm})$ with bosonic
Matsubara frequency $\Omega_{nm}=\omega_n-\omega_m$, the gap function
$\Delta_{s/t}(\mathbf{k}, i\omega_n)$ satisfies
\begin{align}
\begin{split}
\Delta_{s/t}(\mathbf{k}, i\omega_n)
&= -T\sum_{\omega_m} \int \frac{\, d^2 p}{(2\pi)^2}\,
\mathcal{V}_{s/t}\!\left(\mathbf{p}, i\Omega_{nm}\right)
\\
&\times
\frac{\Delta_{s/t}(\mathbf{k-p}, i\omega_m)}
{\omega_m^{2}+\xi_{\mathbf{k-p}}^{2}
+|\Delta_{s/t}(\mathbf{k-p}, i\omega_m)|^{2}} ,
\label{eq:gap_dynamic_main}
\end{split}
\end{align}
where $\omega_m=(2m+1)\pi T$ and $\xi_{\mathbf{k-p}}=\epsilon_{\mathbf{k-p}}-\mu$.
The momentum-transfer structure of the interaction is crucial because it can
promote unconventional, \textit{i.e.} sign-changing anisotropic pairing.
In particular, strong antiferromagnetic fluctuations yield a spin susceptibility $\chi_s(\mathbf{q})$ at low frequencies that is peaked near $\mathbf{q}=(\mathbf{Q},0)$, thereby enhancing pair scattering with momentum transfer $\mathbf{Q}$.
If $\mathbf{Q}$ connects Fermi-surface regions across which a $d$-wave form factor changes sign, this mechanism naturally favors $d$-wave superconductivity.
By contrast, although the charge-fluctuation contribution $U_{\mathrm{eff}}^{2}\chi_c(q)$ enters with the sign required for attraction, its magnitude in our parameter regime is too small to stabilize a BCS-like $s$-wave state.
Consequently, the leading pairing tendency is driven by spin fluctuations in the vicinity of the spin-density-wave instability.

In the static approximation, we set $\mathcal{V}_{s/t}(\mathbf{p}, i\omega_n-i\omega_m)\approx \mathcal{V}_{s/t}(\mathbf{p})$ and $\Delta_{s/t}(\mathbf{k}, i\omega_n)\approx \Delta_{s/t}(\mathbf{k})$.
Carrying out the resulting Matsubara sum then reduces Eq.~\eqref{eq:gap_dynamic_main} to the well-known static, finite-temperature mean-field gap equation reported in Supplemental Material~\cite{SupplementalMaterial},
Sec.~\ref{static gap}.
In this work, we focus on the paramagnetic regime: 
we determine $U_{\mathrm{eff}}$ and solve the corresponding gap equations in both the spin-singlet and spin-triplet channels.
For the numerical solution, we discretize the Brillouin zone by sampling
$\mathbf{k}=(k_x,k_y)$ with $k_{x,y}\in[0,2\pi)$, chosen such that in the
continuum limit the resulting $\mathbf{k}$-sums converge to properly normalized Brillouin-zone integrals.
To capture the leading pairing symmetries, we introduce a set of real form
factors $\{\varphi_\alpha(\mathbf{k})\}$
(Supplemental Material~\cite{SupplementalMaterial},
Sec.~\ref{sec:supp_basis_expansion}) and expand the gap as

\begin{equation}
\Delta(\mathbf{k},i\omega_n)
=\sum_{\alpha}c_\alpha(i\omega_n)\,\varphi_\alpha(\mathbf{k})\,,
\label{eq:gap_basis_expansion}
\end{equation}
with complex coefficients $c_\alpha(i\omega_n)$ determined self-consistently
(see Supplemental Material~\cite{SupplementalMaterial},
Sec.~\ref{sec:supp_fixed_point_solver}).
The dominant pairing symmetry is identified from the largest weight at the
lowest Matsubara frequency, $i\omega_0=i\pi T$. Using the orthonormality condition
\begin{equation}
\int_{\mathrm{BZ}} \frac{d^2k}{(2\pi)^2}\,
\varphi_\alpha^*(\mathbf{k})\varphi_\beta(\mathbf{k})
= \delta_{\alpha\beta},
\label{eq:orthonormality}
\end{equation}
the gap norm at $i\omega_0$ becomes
\begin{equation}
\begin{split}
\|\Delta(i\omega_0)\|^2
&=
\int_{\mathrm{BZ}} \frac{d^2k}{(2\pi)^2}\,
\Delta^\dagger(\mathbf{k},i\omega_0)\Delta(\mathbf{k},i\omega_0) \\
&=
\sum_\alpha |c_\alpha(i\omega_0)|^2.
\label{eq:gap_norm}
\end{split}
\end{equation}
Accordingly, the leading pairing channel is the one with the largest
contribution to $\|\Delta(i\omega_0)\|$.
So this gap amplitude provides a compact, gauge-independent measure of the total superconducting amplitude across all symmetry channels.

We finally emphasize the range of validity of the approximations employed
here. The SRIKR saddle point incorporates the interaction-induced
Gutzwiller renormalization of the quasiparticle background and is not a
weak-coupling expansion in $U$. The subsequent treatment of collective
modes, however, is based on Gaussian fluctuations around the homogeneous
paramagnetic saddle point, while the superconducting interaction is
constructed within a fluctuation-exchange approximation using the
corresponding spin and charge response functions. This construction does
not constitute a systematically controlled expansion with a formal small
parameter over the entire interaction range. It is therefore expected to
be most reliable as long as the paramagnetic saddle point remains locally
stable and the system is not asymptotically close to a collective
instability. Upon softening of a magnetic or charge mode, higher-order
fluctuations, vertex corrections, self-energy feedback, and eventually a
symmetry-broken saddle point may become important. Accordingly, our
superconducting results are interpreted as pairing tendencies of the
correlated paramagnetic metal and are not extrapolated into the
magnetically ordered regime.

\begin{figure*}[t]
    \centering
    \includegraphics[width=0.95\linewidth]{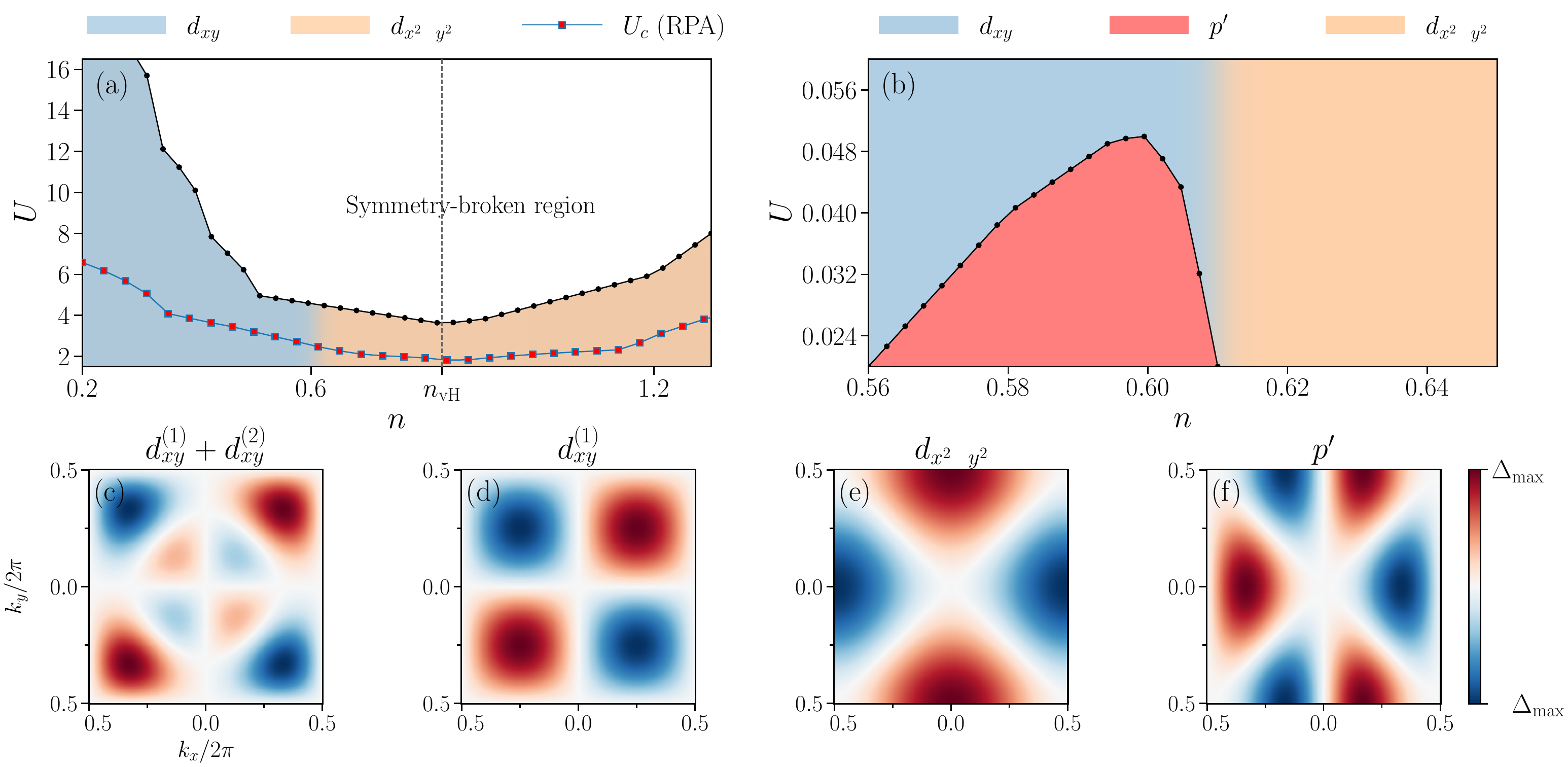}
    \caption{$n$--$U$ phase diagrams at $t'=-0.2$ and $T=0.02\,t$.
(\textbf{a}) Paramagnetic-state superconducting phase diagram, where the
color denotes the leading pairing symmetry. Within the $d_{xy}$ sector,
higher-harmonic contributions are particularly pronounced up to
$n\simeq 0.35$, even at large $U$. With increasing filling, the harmonic
weight is progressively redistributed toward the lowest $d_{xy}^{(1)}$ harmonic,
without change of pairing symmetry. A $d_{xy}\!\to d_{x^2-y^2}$ transition occurs
near $n\simeq 0.61$. The overall phase diagram agrees with
weak-to-intermediate-coupling RPA results
\cite{Romer2020p,Romer2015p} and CPQMC simulations \cite{Cao2025d}.
(\textbf{b}) Low-$U$ phase diagram showing the leading instability,
including the triplet $p'$ phase at $T=0.01\,t$. The $p'$ state belongs to
the $E_u$ irreducible representation and corresponds to a six-node triplet
solution, consistent with Refs.~\cite{Deng2015e,Eremin2002e}.
(\textbf{c})--(\textbf{f}) Representative momentum-space gap structures
for a $d_{xy}$ state with a sizable $d_{xy}^{(2)}$ higher-harmonic
admixture, a state dominated by the lowest $d_{xy}^{(1)}$ harmonic,
$d_{x^2-y^2}$, and $p'$, respectively. Red markers indicate the
RPA-predicted critical interaction $U_c$ in the spin channel, with $U_c$ marking the onset of magnetic order. The
van-Hove filling is marked by a gray dashed line at
$n_{\mathrm{vH}}=0.83$. The symmetry-broken region shown in white corresponds to the magnetically ordered regime, where the paramagnetic saddle point becomes unstable. 
    }
    \label{fig:2}
\end{figure*}

\begin{figure*}[t]
    \centering
    \includegraphics[width=0.95\linewidth]{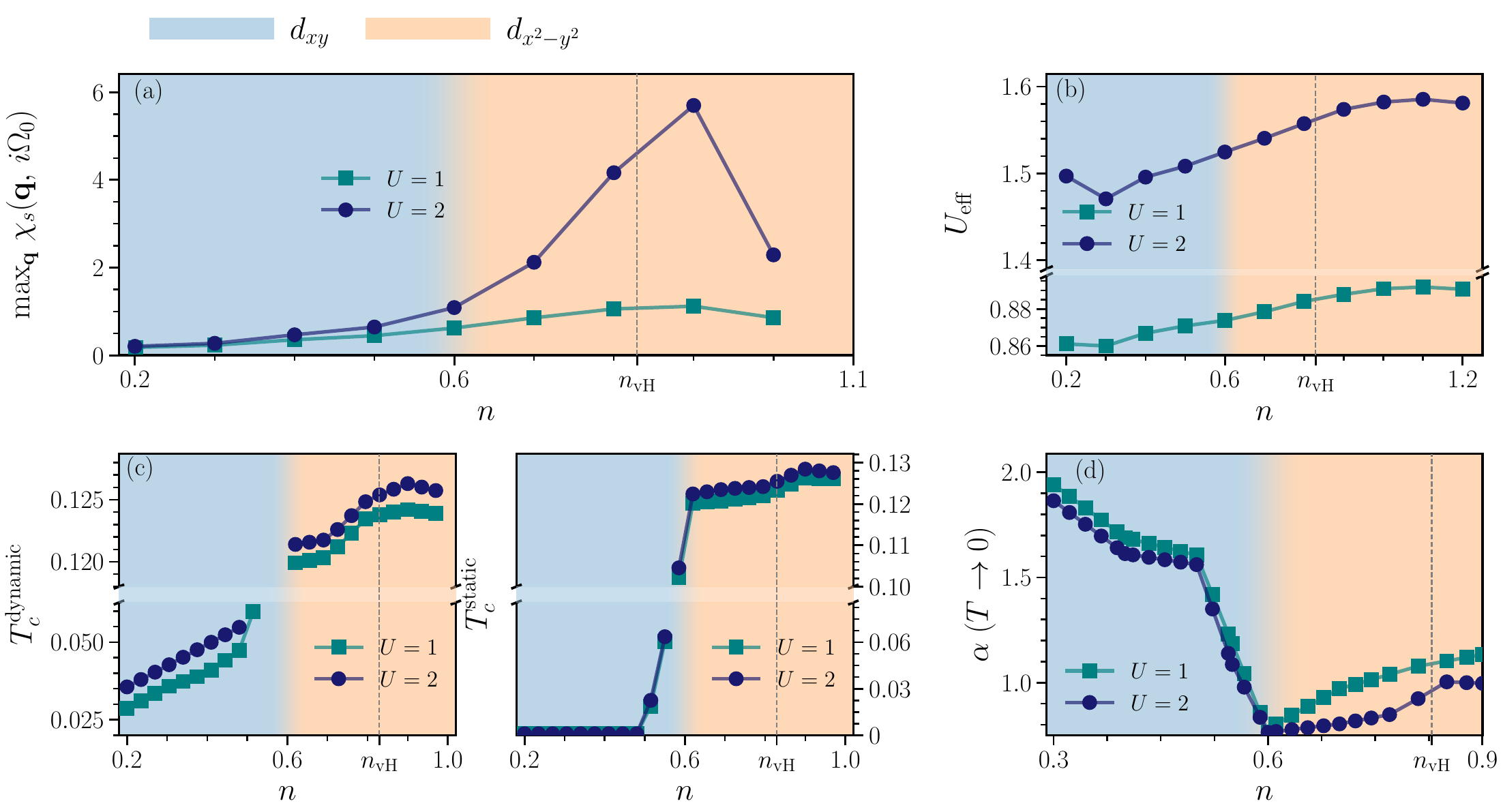}
    \caption{
Filling dependence of magnetic response, effective interaction scale, and superconducting properties. 
(a) Maximum static spin susceptibility $\max_{\mathbf{q}}\chi_s(\mathbf{q},i\Omega_0)$ versus filling $n$ for $U=1$ and $U=2$.
(b)  Effective interaction scale $U_{\mathrm{eff}}$ determined by matching the Stoner-enhanced $\chi_s(\mathbf{Q})$ to an RPA form at the dominant wave vector $\mathbf{Q}$; axis break indicates the different magnitude for $U=1$ and $U=2$. 
(c) Estimated critical temperatures obtained from the full frequency-dependent (dynamic) gap equation (left) and from the static approximation (right);
(d) Exponent $\alpha$ characterizing the decay of the Matsubara-axis gap, $\Delta(i\omega_n)\sim|\omega_n|^{-\alpha}$, shown as a function of $n$.
Background shading indicates the leading singlet symmetry region ($d_{xy}$ and $d_{x^2-y^2}$), and the vertical dashed line marks the van-Hove filling $n_{\textrm{vH}}$. }

    \label{fig:3}
\end{figure*}


\section{Results} \label{sec:re}
\subsection{Spin/charge fluctuation landscape}
We analyze the spin susceptibilities and Fermi surface landscape across the full doping range for a next-nearest-neighbor hopping $t'=-0.2$ motivated by the cuprate Fermiology \cite{Pavarini2001b}, where we mainly focus on the hole-doping.
The zero-frequency limit $\chi_{s/c}(\mathbf{q},i\Omega_0)$ 
provides a compact measure of the low-energy
fluctuation landscape: peaks of $\chi_s(\mathbf{q})$ identify the wave vectors at which magnetic correlations
are strongest and, in spin-fluctuation pairing scenarios, indicate the dominant momentum transfers that enter
the effective pairing interaction \cite{Berk1966e,Scalapino2012a,Chubukov2003a}.
While a fully dynamical treatment is required for quantitative Eliashberg analysis, the static spectrum is
widely used as a transparent approximation and a guide to pairing intuition, because it already captures how
the relevant scattering channels evolve with filling
\cite{Scalapino2012a,Moriya2000s,Monthoux1994sp,Bickers1989c}.

Fig.~\ref{fig:1} summarizes the evolution of the static fluctuation spectrum across the corresponding 
filling range. The first three rows (Fig.~\ref{fig:1}a-\ref{fig:1}i) show the static ($i\Omega_0$) spin susceptibility $\chi_s(\mathbf{q})$, charge $\chi_c(\mathbf{q})$, and the bare susceptibility/Lindhard bubble $\chi_0(\mathbf{q})$, while the last row displays the corresponding Fermi surfaces and the projected nodal features of the leading SC gap, derived by solving the dynamic gap equation as delineated above. At low filling ($n \lesssim 0.4$), both $\chi_s$ and $\chi_c$ display their strongest intensity near the zone boundaries at $(\pm\pi, 0)$ and $(0, \pm\pi)$, consistent with dominant particle--hole scattering across a small electron-like Fermi surface. As $n$ increases, these maxima gradually shift away from the high-symmetry points and evolve into the characteristic quartet of incommensurate peaks at $(\pi \pm \delta, \pi)$ and $(\pi, \pi \pm \delta)$~\cite{Brinckmann1999s,Romer2020p}, signaling the development of increasingly well-nested wave vectors and the approach to nearly perfect nesting as the Fermi surface expands.
Close to half-filling, the response sharpens and becomes strongly concentrated around $\bm q = (\pi, \pi)$, reflecting the enhancement of commensurate antiferromagnetic correlations. Throughout this evolution, the charge channel remains comparatively weak and only weakly structured, indicating that the pairing-relevant interaction is predominantly mediated by spin fluctuations.

This evolution of the Fermi surface underlies the momentum-resolved spin and charge responses discussed above and is shown in the bottom row of Fig.~\ref{fig:1}. For low densities ($n \lesssim 0.4$), the Fermi surface forms a small, nearly circular electron-like pocket centred at $\Gamma$, such that the dominant particle--hole scattering arises from momenta connecting opposite sides of this pocket. With increasing filling, the Fermi surface expands toward the Brillouin-zone boundary and progressively develops flatter segments. At the van-Hove filling (for $t'=-0.2$, this occurs near $n_{\text{vH}} \simeq 0.83$), the Fermi surface touches the saddle points at $(\pm\pi,0)$ and $(0,\pm\pi)$. Beyond this Lifshitz point, the topology becomes hole-like with a large Fermi surface centered around $(\pi,\pi)$; the emergence of extended, nearly parallel segments enhances the phase space for near-nesting scattering and tends to promote incommensurate wave vectors in the spin response.

\subsection{$n$--$U$ Phase diagrams}
\label{phase_diagrams}
We chart the leading superconducting instability as a function of density $n$ and interaction strength $U$ at fixed $t'=-0.2$ and $T=0.02\,t$, using the dynamical fluctuation-mediated pairing vertices of Eqs.~\ref{eq:Vs_main}--\ref{eq:Vt_main}. For each $(n,U)$, we construct $\mathcal{V}_{s/t}(q)$ from the slave-boson susceptibilities, project the nonlinear gap equation onto the chosen form-factor basis, and identify the leading solution from the form factor with the largest low-frequency weight. Since the gap equation is nonlinear, we converge to a unique solution via self-consistent iteration on the projected basis. The intermediate-representation basis and the sparse sampling of imaginary-time and Matsubara-frequency quantities are implemented using the \texttt{sparse-ir} library~\cite{Wallerberger2023s}. For more details on the computational workflow, see Supplemental
Material~\cite{SupplementalMaterial},
Fig.~\ref{fig:computational_workflow}.

Fig.~\ref{fig:2}(a) depicts the singlet superconducting phase diagram,
together with the boundary to the magnetically ordered region, shown in
white. This boundary is taken from the spin-rotation-invariant
slave-boson analyses of Refs.~\cite{Riegler2020s,Seufert2021b}. It marks
the instability of the paramagnetic saddle point toward a static magnetic
mode: upon approaching the boundary from the paramagnetic side, the
static spin susceptibility at the leading ordering vector $\mathbf{Q}$
diverges. The resulting ordered
state may be commensurate or incommensurate, depending on the filling and
interaction strength. Because the pairing vertices employed here are
derived from Gaussian fluctuations around the paramagnetic saddle point,
our superconducting calculation is restricted to the region below this
magnetic-instability boundary. The white region is therefore included
only to indicate where the paramagnetic formulation ceases to apply and
is not assigned a superconducting pairing symmetry.

The superconducting part of the phase diagram contains two symmetry-distinct pairing sectors, with a continuous evolution of the harmonic composition within the low-filling $B_{2g}$ sector. At low filling, up to $n \simeq 0.35$ even for large $U$, the leading instability is a $d_{xy}$ state with sizable higher-harmonic contributions. In this regime, the Fermi surface forms a small, nearly circular pocket around the Brillouin-zone center, while the broad maxima of the spin susceptibility near $(\pm\pi,0)$ and $(0,\pm\pi)$ favor a momentum-space gap structure that is not optimally captured by the lowest harmonic alone.

On such a Fermi surface, the higher $B_{2g}$ lattice harmonics exhibit similar angular profiles, i.e. they share the same nodal structure along the $k_x$ and $k_y$ axes, and are therefore not sharply distinguishable. In particular, the next higher harmonic, $d_{xy}^{(2)}\propto \sin k_x\sin 2k_y+\sin 2k_x\sin k_y$, provides additional flexibility for adapting the gap profile to the momentum dependence of the fluctuation-mediated pairing interaction shown in Fig.~\ref{fig:1}. More generally, lattice harmonics within the same $B_{2g}$ irreducible representation can be constructed systematically.
Although the gap equation is solved within the whole Brillouin zone, our results indicate that, within this formalism, an enhanced variation of the gap along the Fermi surface leads to a more efficient energy gain. Consequently, the low-filling state should be understood as a $d_{xy}$ state with higher-harmonic admixtures, rather than as a distinct superconducting phase. At intermediate fillings, the harmonic weight is progressively redistributed toward the lowest $d_{xy}^{(1)} \propto \sin k_x \sin k_y$ component.

Toward higher filling,
a $d_{x^2-y^2}$ state becomes dominant, with a clear
$d_{xy}\!\to d_{x^2-y^2}$ symmetry transition near $n\simeq0.61$.
The evolution of the dominant pairing symmetry closely follows the structure of the spin susceptibility $\chi_s(\vb{q})$ in Fig.~\ref{fig:1}: when the AF-enhanced wave vectors connect Fermi-surface regions where candidate $d$-wave form factors have opposite signs, those $d$-wave channels are selected, with the competition between $d_{xy}$ and $d_{x^2-y^2}$ determined by the Fermi-surface geometry and the location of the spin-response maxima. This global pattern is quantitatively consistent with weak to intermediate coupling RPA studies and with constrained-path QMC benchmarks, both in terms of the assigned pairing symmetry and in the approximate position of the $d_{xy}$--$d_{x^2-y^2}$ boundary~\cite{Romer2020p,Romer2015p,Cao2025d}. We directly compare our results with the RPA-predicted critical interaction ($U_c$) in the spin channel in Fig.~\ref{fig:2}(a). Since RPA retains only particle–hole ladder diagrams and neglects other scattering channels, it typically underestimates the overall critical scale, particularly at dopings far from the van-Hove singularity \cite{Klett2024i}.
Fig.~\ref{fig:2}(b) reveals a narrow $p'$ region situated in the transition zone between the $d_{xy}$ and $d_{x^2-y^2}$ sectors. At lower $U$, this region coincides with the dominance of the triplet channel.
The $p'$ state belongs to the $E_u$ irreducible representation and is spanned by two degenerate basis functions, of which we show only one; the other is obtained by a 90-degree rotation and has a node along the $k_x$ axis. On the Fermi surface, this state displays a six-node structure, consistent with earlier diagrammatic Monte Carlo and FLEX+RPA results~\cite{Deng2015e,Romer2020p}. Fig.~\ref{fig:2}(c--f) illustrates the representative gap structures in momentum space.
Notably, the $p'$ gap function shares nodes with both neighboring $d$-wave states, making it a consistent intermediate solution at the phase boundary where the transition is near $U=0$. 

We emphasize that the large-$U$ part of the depicted phase diagram should be viewed primarily as theory space for internal consistency checks and for classifying pairing tendencies within the model, rather than as a directly material-motivated regime.

\subsection{Filling dependence of fluctuation \& pairing scale}
\label{filling_dep}
The momentum-resolved susceptibilities in Fig.~\ref{fig:1} can be summarized by two filling-dependent quantities entering the pairing kernel: (i) the overall magnitude of magnetic fluctuations, quantified by the peak value of the static spin susceptibility, $\max_{\mathbf{q}}\chi_s(\mathbf{q},i\Omega_0)$, and (ii) the effective interaction scale $U_{\mathrm{eff}}(n)$ extracted at the dominant magnetic wave vector $\mathbf{Q}$ (Sec.~\ref{vertex}). Although the pairing kernel receives contributions from higher bosonic Matsubara frequencies, its qualitative behavior is largely determined by the lowest bosonic frequency, which typically carries the dominant weight (see Supplemental Material~\cite{SupplementalMaterial},
Fig.~\ref{fig:supp_bosonic}). Fig.~\ref{fig:3}(a,b) depicts both quantities for representative interactions $U=1,2$, with background shading indicating the leading singlet symmetry regions from Fig.~\ref{fig:2}. 

As shown in Fig.~\ref{fig:3}(a), the magnetic response is weak at low filling ($n\lesssim 0.35$), where the system lies in the $d_{xy}$ sector
with sizable higher-harmonic contributions. $\chi_s$ remains small and only weakly varying with $n$, in this regime. As $n$ increases,
the harmonic weight is progressively redistributed toward the lowest
$d_{xy}$ component, while $\max_{\mathbf{q}}\chi_s$ grows more noticeably, consistent with the development of stronger near-nesting scattering channels and the progressive sharpening of the spin response seen in Fig.~\ref{fig:1}. The most pronounced enhancement occurs in the $d_{x^2-y^2}$ sector ($n\gtrsim 0.61$): for $U=2$ the maximal static susceptibility rises rapidly and forms a dome near the high-filling end, signalling proximity to a Stoner-enhanced antiferromagnetic instability. For $U=1$ the increase is more modest but follows the same monotonic trend, indicating that the strengthening of magnetic fluctuations with filling is a robust feature even at weaker couplings. The vertical dashed line marks the van-Hove filling $n_{\mathrm{vH}}$; across this point, the magnetic response continues to grow, consistent with an enhanced phase space for particle--hole scattering and the associated amplification of spin fluctuations.

The extracted $U_{\mathrm{eff}}(n)$ in Fig.~\ref{fig:3}(b) provides a compact interaction scale that enters the fluctuation-exchange vertices of Eqs.~\eqref{eq:Vt_main}--\eqref{eq:Vs_main}. By construction, $U_{\mathrm{eff}}$ is determined from the ratio between the interacting and bare responses at the dominant wave vector $\mathbf{Q}$, and therefore tracks the degree of Stoner enhancement encoded in $\chi_s(\mathbf{Q})$. For both $U=1$ and $U=2$, $U_{\mathrm{eff}}$ increases smoothly with filling, reflecting the systematic strengthening of magnetic correlations as the system moves toward the high-filling $d_{x^2-y^2}$ regime. We find that $U_{\mathrm{eff}}$ scales approximately linearly with $U$ (see Supplemental Material~\cite{SupplementalMaterial},
Fig.~\ref{fig:Ueff_U}), and that its renormalization with filling is stronger at larger $U$. Together, Figs.~\ref{fig:3}(a) and~\ref{fig:3}(b) therefore quantify the two central ingredients for pairing: the buildup of spin-fluctuation spectral weight and the resulting enhancement of the effective interaction scale that sets the overall strength of the pairing vertex. Below, we relate these trends to the evolution of the dynamical gap structure and the resulting $T_c$.

\subsection{Critical Temperature $T_c$}
\label{critical_temp}

We extract the superconducting critical temperature $T_c$ from temperature sweeps of the fully self-consistent solutions of the gap equation, using the gap amplitude defined by the Brillouin-zone norm in Eq.~\eqref{eq:gap_norm}; $T_c$ is identified as the highest temperature at which a nontrivial solution persists. Fig.~\ref{fig:3}(c) compares $T_c$ obtained from the full frequency-dependent (dynamic) gap equation, $T_c^{\rm dynamic}$, with the static approximation, $T_c^{\rm static}$, for representative interactions $U=1,2$. In the high-filling $d_{x^2-y^2}$ regime, where the magnetic response is strongest, $T_c^{\rm static}$ systematically exceeds $T_c^{\rm dynamic}$: treating the pairing kernel as instantaneous effectively weights the interaction by its $\Omega=0$ component over the full frequency spectrum and therefore overestimates the pairing strength, while the dynamic equation includes the full $\Omega$ dependence of the spin/charge fluctuations entering the vertex, which reduces the net pairing efficiency and suppresses $T_c$ through retardation and fluctuation-induced scattering encoded in $\mathcal{V}_{s/t}(\mathbf{p},i\Omega_{nm})$.

The doping dependence of $T_c$ closely tracks the evolution of the spin fluctuations: $\max_{\mathbf{q}}\chi_s(\mathbf{q},i\Omega_0)$ grows strongly upon approaching the $d_{x^2-y^2}$ region [Fig.~\ref{fig:3}(a)], and at the same time both $U_{\rm eff}$ [Fig.~\ref{fig:3}(b)] and $T_c$ are enhanced, consistent with spin-fluctuation-dominated pairing~\cite{Kitatani2015f}.
By contrast, at low fillings, \textit{i.e.} in the $d_{xy}$ regime, the spin response is weaker and broader in momentum space, yielding substantially smaller $T_c$ in both the static and dynamic treatments. Intuitively, the dynamic transition typically exceeds the static one, reflecting the fact that the full frequency dependence of the pairing kernel allows pairing to persist at higher temperatures than the zero-frequency approximation, even when the static Stoner enhancement is weak. In this regime, the reduced Stoner enhancement and the less favorable
projection of the broad static response onto the $d_{xy}$ gap structure
limit the effective pairing scale in the static approximation. By
contrast, the dynamical equation accesses finite-frequency spectral weight
with a more favorable momentum-space overlap and additionally couples
different fermionic Matsubara frequencies, thereby supporting a finite
solution at a somewhat higher $T_c$. This is consistent with experimental expectations in cuprate systems, where accessible superconductivity emerges primarily at higher fillings and not at the low-filling end of the phase diagram; the strong suppression of $T_c$ at low fillings indicates that unconventional superconductivity is unlikely to be observed in this regime, given typical experimental sensitivity and sample‑quality requirements.

Notably, the estimated $T_c(n)$ peaks away from the van-Hove filling $n_{\rm vH}$ (Fig.~\ref{fig:3}(c)), qualitatively aligning with cuprate experiments where optimal hole-doping precedes the van-Hove singularity \cite{Benhabib2015c,Piriou2011f,Horio2018t,Storey2007s}, though our generic model eschews material-specific predictions. As we will point out in the following sections the real-frequency gap $\mathfrak{R}\Delta(\omega)$ exhibits a dynamical softening with increasing $n$: a pronounced low-energy peak structure $\omega_{\rm peak}(n)$ reduces with increasing filling. Such a feature naturally identifies the characteristic (retarded) scale of the effective pairing glue provided by spin fluctuations. In realistic materials, stronger retardation may amplify this feedback, raising $T_c$ beyond our RPA-like estimates \cite{Monthoux1994se,Carbotte1990p}. 

\subsection{Frequency dependent Gap profile}
\label{gap_dynamics}

Matsubara-axis solutions of the full (dynamic) gap equation provide direct access to the frequency structure of the superconducting order parameter. The analysis starts from discrete data $\{z_n=i\omega_n,\Delta(i\omega_n)\}$ obtained from the self-consistent solver, where $\Delta(i\omega_n)$ is evaluated at a representative momentum on the Fermi surface. The large-$|\omega_n|$ tail is well described by a power law, $\Delta(i\omega_n)\sim |\omega_n|^{-\alpha}$ so that the exponent $\alpha$ provides a compact characterization of the retardation contained in the fluctuation-mediated pairing kernel. 

\begin{figure*}[t]
    \centering
    \includegraphics[width=\textwidth]{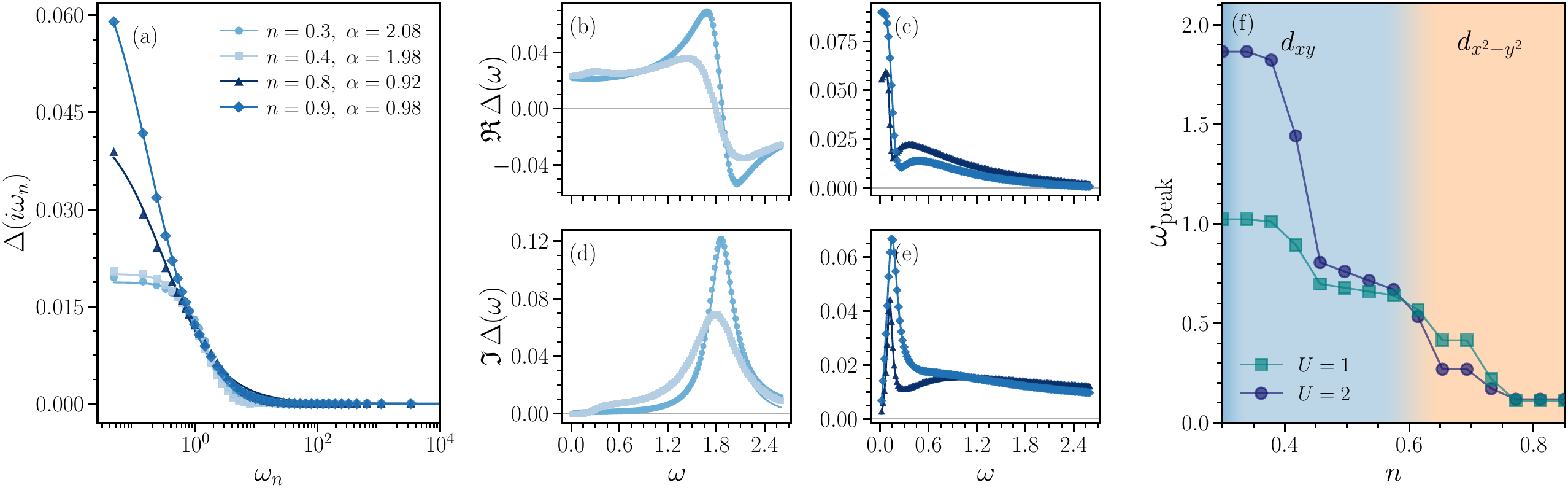}
    \caption{
    Frequency dependence and real-axis structure of the dynamical
    superconducting gap for $t'=-0.2$.
    (a) Matsubara-frequency dependence of the leading gap
    $\Delta(i\omega_n)$ for $U=2$ at representative fillings.
    Symbols denote the numerical results and dashed lines show power-law fits. The larger values of
    $\alpha$ in the low-filling $d_{xy}$ regime indicate a more rapidly
    decaying frequency dependence, whereas $\alpha\simeq1$ toward the
    $d_{x^2-y^2}$ regime is consistent with pairing dominated by strong
    antiferromagnetic fluctuations.
    (b)--(e) Pad\'e analytic continuation of the dynamical gap for
    $U=2$. Panels (b) and (c) show
    $\mathfrak{R}\Delta(\omega)$, while panels (d) and (e) show
    the corresponding $\mathfrak{I}\Delta(\omega)$ for the
    low- and high-filling regimes, respectively. Symbols denote the
    Pad\'e-reconstructed data, and solid lines show the fit to the real
    part and the corresponding Kramers--Kronig result for the imaginary
    part.
    (f) Peak position $\omega_{\mathrm{peak}}$ of
    $\mathfrak{R}\Delta(\omega)$, extracted using the fit form in
    Eq.~\eqref{eq:DeltaRfit_Re}, as a function of filling for $U=1$ and
    $U=2$. The background shading indicates the leading singlet pairing
    regions from Fig.~\ref{fig:2}. With increasing filling,
    $\omega_{\mathrm{peak}}$ shifts systematically toward lower
    frequencies and exhibits a pronounced change across the
    $d_{xy}\!\to d_{x^2-y^2}$ symmetry transition.
    }
    \label{fig:4}
\end{figure*}

The filling dependence of $\alpha$ is summarized in Fig.~\ref{fig:3}(d). In the $d_{x^2-y^2}$ regime, values $\alpha\sim 1$ indicate a slowly decaying gap tail, consistent with pairing dominated by strong antiferromagnetic fluctuations. By contrast, the larger $\alpha$ values in the low-density $d_{xy}$ sector correspond to a more rapidly decaying gap response, suggesting a less strongly retarded effective pairing kernel [Fig.~\ref{fig:4}(a)]. A pronounced reduction of $\alpha$ occurs near the symmetry-changing boundary, signalling enhanced low-frequency pairing fluctuations associated with competition between nearly degenerate $d_{xy}$ and $d_{x^2-y^2}$ channels. This behavior is naturally connected to quantum-critical pairing scenarios such as Chubukov's $\gamma$-model, where a singular bosonic glue $\lambda(\Omega_m)\propto 1/|\Omega_m|^{\gamma}$ can generate power-law Matsubara-frequency structures over an extended scaling window
\cite{Wang2016s,Wu2019s}. The same dynamical structure is also reflected in the systematic suppression of the dynamic critical temperature relative to the static approximation towards high-filling, discussed in Fig.~\ref{fig:3}(c).

While $\Delta(i\omega_n)$ is the natural output of the dynamical gap equation, identifying characteristic pairing scales requires access to the real-frequency gap $\Delta(\omega)$. To characterize the frequency dependence of the full gap function
$\Delta(\mathbf{k},i\omega_n)$, we evaluate it at a representative
non-nodal momentum $\mathbf{k}_{\star}$ on the interacting Fermi surface.
For each filling, $\mathbf{k}_{\star}$ is chosen as the point at which
$|\Delta(\mathbf{k},i\omega_0)|$ is maximal along the Fermi surface. The complete
momentum dependence remains encoded in the frequency-dependent
form-factor coefficients of
Eq.~\eqref{eq:gap_basis_expansion}, so that different lattice harmonics
may retain distinct frequency dependences within the chosen basis. Notably, some studies suggest \cite{Manske2003r,Abanov2003q} that there can be a nontrivial intertwining of the functional spaces in $(\mathbf{k},\omega)$, but in our case such coupling appears to be weak. Methodologically, however, our approach does not hinge on the product ansatz and can be generalized beyond it.

Because analytic continuation is numerically demanding and not generally feasible, we construct an analytic approximation $\Delta(z)$ from the Matsubara-axis data using an $N$-point Pad\'e approximant $P_N(z)$ in the continued-fraction form~\cite{Vidberg1977s,Baker1975e} (see Supplemental Material~\cite{SupplementalMaterial},
Sec.~\ref{sec:supp_freq_gap}),
\begin{equation}
\Delta(\omega)=\lim_{\eta\to 0^+}P_N(\omega+i\eta).
\end{equation}

Analytic continuation is here more straightforward: for $\omega>0$, the resulting $\mathfrak{R}\Delta(\omega)$ is well described by a minimal phenomenological form that captures both a pronounced finite-frequency structure around a characteristic scale $\omega_p$ and a slowly decaying high-frequency tail,
\begin{equation}
\mathfrak{R}\Delta(\omega)=
c_0
+ C\,\frac{\omega-\omega_p}{(\omega-\omega_p)^2+\Gamma^2}
+ \frac{1}{A+B\omega^{\lambda}}.
\label{eq:DeltaRfit_Re}
\end{equation}
The constant offset $c_0$ is retained to improve the numerical stability of the fit and to absorb small residual offsets from the Pad\'e continuation; in practice, however, we find $c_0$ to be negligible in the subsequent analysis. The second term is dispersive in the sense that it describes a resonant, frequency-dependent correction with a non-constant spectral shape. On its own, this term corresponds to a Lorentzian-like response in the time domain, reflecting the finite lifetime and oscillatory decay of the underlying fluctuation mode. The last term reproduces the algebraic tail already visible on the Matsubara axis. The corresponding fits for both $\mathfrak{R}\Delta(\omega)$ and $\mathfrak{I}\Delta(\omega)$ are shown in Fig.~\ref{fig:4} (b-e). The imaginary part is not fitted independently. Instead, it is constrained by causality and obtained from the Kramers--Kronig relation,
\begin{align}
\mathfrak{I}\Delta(\omega)
&=
-\frac{1}{\pi}\,\mathcal{P}\!\!\int_{-\infty}^{\infty}
\frac{\mathfrak{R}\Delta(\omega')}{\omega'-\omega}\,d\omega',
\label{eq:KK_general_Im}
\\
\mathfrak{R}\Delta(\omega)
&=
\frac{1}{\pi}\,\mathcal{P}\!\!\int_{-\infty}^{\infty}
\frac{\mathfrak{I}\Delta(\omega')}{\omega'-\omega}\,d\omega'.
\label{eq:KK_general_Re}
\end{align}
This procedure yields excellent agreement for $\mathfrak{I}\Delta(\omega)$ and provides a stringent consistency check on the real-axis continuation. The fit also makes the limiting behavior transparent:
\begin{equation}
\mathfrak{R}\Delta(\omega\to 0)=
- C\,\frac{\omega_p}{\omega_p^2+\Gamma^2}
+ \frac{1}{A},
\label{eq:loww_limit}
\end{equation}
At low filling, the real-frequency gap function is well described by a purely dispersive form, whereas at higher filling, an additional broad tail is required to accurately capture the data over the accessible frequency window. The dispersive part decays as $1/\omega$ for large $\omega$, while the tail contributes a correction $\sim \omega^{-\lambda}$. For the stable high-filling fits we find $\lambda>1$, implying that this tail is subleading in the $\omega\to\infty$ limit. Our parametrization captures both the characteristic intermediate-frequency structure and the subleading power-law correction visible in Fig.~\ref{fig:4}(a), while leaving the leading asymptotic behavior consistent. This interpretation is consistent with earlier studies reporting enhanced finite-frequency structure in dynamical gap functions, although without showing an explicit fitting framework \cite{Monthoux1994se,Vidberg1977s,Dahm1995q,Wu2019s,Gao2024r}.


Upon Pad\'e continuation, $\mathfrak{R}\Delta(\omega)$ develops a pronounced finite-frequency peak whose position $\omega_{\rm peak}(n)$ and magnitude evolve systematically with filling. We extract $\omega_{\rm peak}$ from the fit in Eq.~\eqref{eq:DeltaRfit_Re} and find the same qualitative trend when comparing the slave-boson results to FLEX+RPA at the same $(U,t')$ (see Supplemental Material~\cite{SupplementalMaterial},
Fig.~\ref{fig:5}). As $n$ increases, $\omega_{\rm peak}$ shifts to lower frequencies and exhibits a clear change in slope near the $d_{xy}\!\to d_{x^2-y^2}$ crossover [Fig.~\ref{fig:4}(f)]. Since the momentum dependence is already encoded in the form-factor expansion, this trend reflects how the leading pairing channel samples different parts of the spin-fluctuation spectrum as the filling is varied. In the language of spin-fluctuation pairing, the downward shift of $\omega_{\rm peak}$ indicates a softening of the characteristic energy scale of the effective pairing interaction, i.e. the dominant spectral weight of the pairing glue moves to lower frequencies near half filling \cite{Bickers1989c,Monthoux1994se,Manske2003r,Abanov2003q}. The overall trend is broadly consistent with experiment, in Bi$_2$Sr$_2$CaCu$_2$O$_{8+\delta}$, high-resolution ARPES revealed clear superconductivity-induced changes in the near-nodal electronic excitations \cite{Valla2006f}. Subsequent measurements on overdoped Bi2212 showed that the renormalization weakens with doping and disappears together with superconductivity, identifying the onset of the spin-fluctuation spectrum as the relevant low-energy scale \cite{Valla2020d}. In this sense, the reduction of $\omega_{\rm peak}$ with increasing filling in our calculation can be viewed as a signature of the same softening tendency in the characteristic electronic renormalization scale \cite{Fong2000s,Brehm2010t}.


Although the real-frequency gap function $\Delta(\mathbf{k},\omega)$ is not directly observable, its frequency dependence enters the anomalous electronic self-energy and therefore influences experimentally accessible spectral functions. A pronounced feature in $\Delta(\omega)$ at the characteristic spin-fluctuation energy scale may give rise to dispersion anomalies or peak--dip--hump structures in ARPES, dip--hump features in tunnelling spectra, and corresponding signatures in the optical self-energy or frequency-dependent scattering rate \cite{McMillan1965l,Carbotte1990p,Eschrig2003e,Zhu2004e,Lee2006i,Yang2009e}. The filling dependence of the peak position found here could thus manifest itself as a systematic carrier-density dependence of these spectroscopic energy scales. This correspondence is necessarily qualitative, however, because the measured responses depend on both the normal and anomalous self-energies, as well as on momentum-dependent matrix elements and quasi-particle broadening. A direct comparison with experiment would therefore require calculating the full electronic spectral function, tunneling conductance, or optical conductivity.

\section{Conclusion}\label{sec:co}


We have developed a superconducting fluctuation-exchange framework within
the spin-rotation-invariant Kotliar--Ruckenstein slave-boson formulation of
the two-dimensional Hubbard model with $t$--$t'$ hopping. Starting from
Gaussian fluctuations around the paramagnetic saddle point, we obtained
fully dynamical spin and charge susceptibilities on top of a correlated,
Gutzwiller-renormalized quasiparticle background. This correlated starting
point distinguishes the present approach from conventional weak-coupling
RPA and FLEX-based pairing schemes, in which the fluctuation-mediated
interaction is constructed from bare or only perturbatively renormalized
quasiparticles. Here, strong-correlation effects enter self-consistently
already through the quasiparticle dispersion, quasiparticle weight, and
collective response functions. The resulting susceptibilities were used to
construct an effective two-particle pairing vertex that is explicitly
antisymmetric under fermionic exchange and can be projected onto singlet
and triplet channels. Solving the resulting anisotropic and
frequency-dependent gap equation in a form-factor basis allowed us to map
the leading superconducting instabilities as functions of filling and
interaction strength and to resolve both the Matsubara- and real-frequency
structure of the superconducting gap. The framework therefore combines
local-correlation-induced quasiparticle renormalization, dynamical
collective fluctuations, and superconducting pairing within a single
consistent formalism, while recovering the established filling dependence
of the dominant pairing symmetries obtained in earlier RPA and FLEX studies.

The central physical result is that pairing is dominated throughout by the spin-fluctuation channel. In the paramagnetic regime studied here, the leading singlet symmetry evolves systematically from a low-filling $d_{xy}$ regime with higher-harmonics admixture to an intermediate $d_{xy}^{(1)}$ regime and finally to a high-filling $d_{x^2-y^2}$ state. This sequence closely follows the filling-dependent evolution of $\chi_s(\mathbf{q})$ and the corresponding dominant scattering wave vectors, consistent with the standard sign-changing spin-fluctuation mechanism. At the same time, both the maximal static spin response and the interaction scale $U_{\rm eff}$ extracted from the Stoner enhancement grow toward higher filling, leading to an enhancement of the pairing scale and of the estimated $T_c$. While the low-filling sectors are useful for classifying the intrinsic pairing tendencies of the model theoretically, the higher-filling $d_{x^2-y^2}$ regime connects more directly to the parameter range usually associated with cuprate phenomenology.

The dynamical gap structure provides additional information beyond the symmetry classification. On the Matsubara axis, the gap shows a robust power-law decay $\Delta(i\omega_n)\sim |\omega_n|^{-\alpha}$ with a filling-dependent exponent, and the fully dynamic calculation yields systematically lower critical temperatures than the static approximation for the high-filling regime, demonstrating the importance of retardation effects. After Pad\'e continuation, the real part of the gap develops a pronounced finite-frequency peak whose position $\omega_{\rm peak}$ shifts to lower energies with increasing filling and changes slope near the $d_{xy}$ to $d_{x^2-y^2}$ crossover. We interpret this as a dynamical fingerprint of the softening of the characteristic pairing scale encoded in the spin-fluctuation spectrum. In this sense, the real-frequency analysis provides a direct way to characterize how the effective pairing glue evolves across the phase diagram.

Several extensions of the present approach suggest themselves and provide a roadmap for future work. Superconducting phases emerging on top of
spin- or charge-ordered states should become accessible once the present
methodology is extended to calculate the corresponding collective
susceptibilities around symmetry-broken slave-boson saddle points. On the methodological side, it would be valuable to incorporate additional self-energy feedback beyond the current quasiparticle renormalization, to capture the back-action of superconductivity on the fluctuation spectrum and to allow for competing symmetry-broken states beyond the paramagnetic saddle, such as magnetic or charge order. Extending the framework to multi-orbital settings would further pave the way toward more material-specific applications. More broadly, the present work establishes a scalable route from slave-boson fluctuation dynamics to microscopic pairing interactions, superconducting phase competition, and dynamical gap properties in correlated lattice systems, which can serve as a basis for such future refinements.

\section{Acknowledgement}

 We thank T. Saha-Dasgupta, I. Dasgupta, M.~Dürrnagel, L.~Klebl and T.~Müller for fruitful discussions. 
This research was funded by the Deutsche Forschungsgemeinschaft (DFG, German Research Foundation) – Project-ID 258499086 – SFB 1170; through the Würzburg-Dresden Cluster of Excellence on Complexity, Topology and Dynamics in Quantum Materials (ctd.qmat) – Project-ID 390858490 – EXC 2147; P. W\"olfle acknowledges support through a Distinguished Senior Fellowship of Karlsruhe Institute of Technology.

\let\oldaddcontentsline\addcontentsline
\renewcommand{\addcontentsline}[3]{}
\bibliography{ref}
\let\addcontentsline\oldaddcontentsline

\clearpage

\supplement{Supplemental Material for\\
Slave-boson Formalism for Superconducting Pairing at Strong Coupling}
\onecolumngrid
\suppTOCon
\setcounter{tocdepth}{2}
\supplementarytableofcontents

\section{Slave-boson observables}
\label{sec:sb}

For completeness, we summarize the spin-rotation-invariant Kotliar--Ruckenstein
(SRIKR) slave-boson formulation used in the main text. We focus on the definitions
needed to identify the charge and spin observables entering the Gaussian
fluctuation analysis.
\subsection{SRIKR representation}

At each lattice site \(i\), the enlarged local Hilbert space is generated by
pseudofermions \(f_{i\sigma}\) and slave bosons \(e_i,d_i,p_{i\mu}\), with
\(\mu=0,1,2,3\). The bosons \(e_i\) and \(d_i\) describe the empty and doubly
occupied states, while the spin-rotation-invariant singly occupied sector is
encoded in the \(2\times2\) matrix
\begin{equation}
p^\dagger_i
=
\frac{1}{2}\sum_{\mu=0}^{3}p^\dagger_{i\mu}\tau^\mu
=
\frac{1}{2}
\begin{pmatrix}
p^\dagger_{i0}+p^\dagger_{i3} & p^\dagger_{i1}-i p^\dagger_{i2} \\
p^\dagger_{i1}+i p^\dagger_{i2} & p^\dagger_{i0}-p^\dagger_{i3}
\end{pmatrix},
\label{eq:supp_p_matrix}
\end{equation}
where \(\tau^0\) is the \(2\times2\) identity matrix and
\(\tau^{1,2,3}\) are Pauli matrices. The physical local states are represented as
\begin{equation}
\ket{0}_i
=
e_i^\dagger\ket{\mathrm{vac}},
\qquad
\ket{\sigma}_i
=
\sum_{\sigma'}
p^\dagger_{i,\sigma\sigma'}
f^\dagger_{i\sigma'}\ket{\mathrm{vac}},
\qquad
\ket{2}_i
=
d_i^\dagger
f^\dagger_{i\uparrow}f^\dagger_{i\downarrow}
\ket{\mathrm{vac}}.
\label{eq:supp_local_states}
\end{equation}
The physical Hilbert space is selected by the local constraints
\begin{subequations}
\label{eq:supp_sb_constraints}
\begin{align}
1
&=
e_i^\dagger e_i
+d_i^\dagger d_i
+\sum_{\mu=0}^{3}p^\dagger_{i\mu}p^\nodag_{i\mu},
\label{eq:supp_constraint_completeness}
\\
\sum_{\sigma}f^\dagger_{i\sigma}f^\nodag_{i\sigma}
&=
\sum_{\mu=0}^{3}p^\dagger_{i\mu}p^\nodag_{i\mu}
+2d_i^\dagger d_i,
\label{eq:supp_constraint_charge}
\\
\sum_{\sigma\sigma'}
\boldsymbol{\tau}_{\sigma\sigma'}
f^\dagger_{i\sigma'}f^\nodag_{i\sigma}
&=
p^\dagger_{i0}\mathbf p_i
+\mathbf p_i^\dagger p^\nodag_{i0}
-i\,\mathbf p_i^\dagger\times\mathbf p_i .
\label{eq:supp_constraint_spin}
\end{align}
\end{subequations}
Here \(\mathbf p_i=(p_{i1},p_{i2},p_{i3})^{\mathrm T}\). The physical electron operator is expressed in terms of pseudofermions as
\begin{equation}
c^\dagger_{i\sigma}
=
\sum_{\sigma'}
z^\dagger_{i,\sigma\sigma'}
f^\dagger_{i\sigma'},
\qquad
c^\nodag_{i\sigma}
=
\sum_{\sigma'}
f^\nodag_{i\sigma'}
z^\nodag_{i,\sigma'\sigma}.
\label{eq:supp_sb_cmap}
\end{equation}
Following the original Kotliar--Ruckenstein operator ordering, the
spin-rotation-invariant renormalization matrix is taken as
\begin{equation}
z_i
=
\sqrt{2}\,
L_i
\left(
e_i^\dagger p_i+\tilde p_i^\dagger d_i^\nodag
\right)
R_i ,
\label{eq:supp_z_factor}
\end{equation}
with
\begin{equation}
z_i^\dagger
=
\sqrt{2}\,
R_i
\left(
p_i^\dagger e_i^\nodag
+d_i^\dagger\tilde p_i^\nodag
\right)
L_i .
\label{eq:supp_z_factor_dagger}
\end{equation}
The matrices \(p_i\) and \(\tilde p_i\) are defined by
\begin{equation}
p_i
=
\frac{1}{2}
\left(
p_{i0}\tau^0+\sum_{a=1}^{3}p_{ia}\tau^a
\right),
\qquad
\tilde p_i
=
\frac{1}{2}
\left(
p_{i0}\tau^0-\sum_{a=1}^{3}p_{ia}\tau^a
\right).
\label{eq:supp_p_tildep}
\end{equation}
The left and right normalization matrices are
\begin{equation}
L_i
=
\left[
\left(1-d_i^\dagger d_i^\nodag\right)\tau^0
-2p_i^\dagger p_i^\nodag
\right]^{-1/2},
\qquad
R_i
=
\left[
\left(1-e_i^\dagger e_i^\nodag\right)\tau^0
-2\tilde p_i^\dagger \tilde p_i^\nodag
\right]^{-1/2}.
\label{eq:supp_LR_factors}
\end{equation}
All products in Eqs.~\eqref{eq:supp_z_factor}--\eqref{eq:supp_LR_factors}
are matrix products in spin space, and the inverse square roots are understood
as matrix inverse square roots. The normalization factors \(L_i\) and \(R_i\)
ensure the correct noninteracting limit and reproduce the Gutzwiller
quasiparticle renormalization at the saddle point.

\subsection{Hamiltonian and effective action}

We start from the one-band Hubbard model
\begin{equation}
H
=
\sum_{ij,\sigma}c^\dagger_{i\sigma}t_{ij}c^\nodag_{j\sigma}
-\mu\sum_i \hat n_i
+
U\sum_i
c^\dagger_{i\uparrow}c^\nodag_{i\uparrow}
c^\dagger_{i\downarrow}c^\nodag_{i\downarrow}.
\label{eq:supp_hubbard_real}
\end{equation}
After the SRIKR substitution, the interacting term becomes \(U d_i^\dagger d_i\),
while the kinetic term is renormalized by the \(z\)-matrices. In momentum space,
the resulting fermionic Hamiltonian can be written as
\begin{equation}
H_{\mathrm{SB}}
=
\sum_{\mathbf k_1,\mathbf k_2}
\mathbf f^\dagger_{\mathbf k_1}
H_{\mathbf k_1,\mathbf k_2}[\psi]
\mathbf f^\nodag_{\mathbf k_2}
+
U\sum_i d_i^\dagger d_i^\nodag,
\label{eq:supp_sb_hamiltonian}
\end{equation}
with \(\mathbf f_{\mathbf k}=(f_{\mathbf k\uparrow},f_{\mathbf k\downarrow})^{\mathrm T}\) and
\begin{equation}
H_{\mathbf k_1,\mathbf k_2}[\psi]
=
-\mu\,\delta_{\mathbf k_1,\mathbf k_2}\mathbbm{1}_2
+
\frac{1}{N}
\sum_{\mathbf k}
\left(z^\dagger\right)^{\mathrm T}_{\mathbf k-\mathbf k_1}
\mathcal H_{\mathbf k}
\left(z^\nodag\right)^{\mathrm T}_{\mathbf k-\mathbf k_2}.
\label{eq:supp_sb_Hk}
\end{equation}
The collective bosonic field is denoted by
\begin{equation}
\psi=(e,d,p_0,\mathbf p,\alpha,\beta_0,\boldsymbol{\beta}),
\label{eq:supp_collective_field}
\end{equation}
where \(\alpha\), \(\beta_0\), and \(\boldsymbol{\beta}\) are Lagrange multipliers
enforcing the constraints in Eq.~\eqref{eq:supp_sb_constraints}. The partition function is written as a coherent-state path integral,
\begin{equation}
Z=
\int\mathcal D[f^*,f]\,\mathcal D[\psi^*,\psi]\,
e^{-\mathcal S[f,\psi]},
\qquad
\mathcal S=\int_0^{1/T}d\tau\,
\left(\mathcal L_{\mathrm B}+\mathcal L_{\mathrm F}\right).
\label{eq:supp_partition_function}
\end{equation}
The bosonic and fermionic parts of the Euclidean Lagrangian are
\begin{subequations}
\label{eq:supp_lagrangian}
\begin{align}
\mathcal L_{\mathrm B}
&=
\sum_i
\Big[
d_i^*(\partial_\tau+U)d_i
+e_i^*\partial_\tau e_i
+p_{0,i}^*\partial_\tau p_{0,i}
+\mathbf p_i^*\!\cdot\!\partial_\tau\mathbf p_i
\nonumber\\
&\quad
+i\alpha_i
\left(e_i^*e_i+p_{0,i}^*p_{0,i}
+\mathbf p_i^*\!\cdot\!\mathbf p_i+d_i^*d_i-1\right)
\nonumber\\
&\quad
-i\beta_{0,i}
\left(p_{0,i}^*p_{0,i}
+\mathbf p_i^*\!\cdot\!\mathbf p_i+2d_i^*d_i\right)
\nonumber\\
&\quad
-i\boldsymbol{\beta}_i\!\cdot\!
\left(
p_{0,i}^*\mathbf p_i+\mathbf p_i^*p_{0,i}
-i\,\mathbf p_i^*\times\mathbf p_i
\right)
\Big],
\\
\mathcal L_{\mathrm F}
&=
\sum_{\mathbf k_1,\mathbf k_2}
\mathbf f^\dagger_{\mathbf k_1}
\left[
\partial_\tau\delta_{\mathbf k_1,\mathbf k_2}
+
H_{\mathbf k_1,\mathbf k_2}[\psi]
\right]
\mathbf f^\nodag_{\mathbf k_2}.
\end{align}
\end{subequations}
Integrating out the pseudofermions gives the bosonic effective action
\begin{equation}
\mathcal S_{\mathrm{eff}}[\psi]
=
\int_0^{1/T}d\tau\,\mathcal L_{\mathrm B}
-
\mathrm{Tr}\ln
\left[
\partial_\tau+H[\psi]
\right],
\label{eq:supp_effective_action}
\end{equation}
which is the starting point for the saddle-point and Gaussian fluctuation
analysis. After a local \(SU(2)\times U(1)\times U(1)\) gauge transformation, we work in
the radial gauge. In this gauge, the phases of \(e\), \(p_0\), and
\(\mathbf p\) are gauged away, so that these fields can be chosen real, while
the doublon field \(d\) remains complex. The corresponding gauge fields appear
as the Lagrange multipliers \(\alpha\), \(\beta_0\), and
\(\boldsymbol{\beta}\), which enforce the local SRIKR constraints. The fermionic
kernel then acquires the local constraint fields explicitly,
\begin{equation}
H_{\mathbf k_1,\mathbf k_2}[\psi]
=
-\mu\,\mathbbm{1}_2\delta_{\mathbf k_1,\mathbf k_2}
+
\frac{1}{\sqrt N}
\left(\beta_0\tau^0+\boldsymbol{\beta}\cdot\boldsymbol{\tau}\right)^{\mathrm T}_{\mathbf k_1-\mathbf k_2}
+
\frac{1}{N}
\sum_{\mathbf k}
\left(z^\dagger\right)^{\mathrm T}_{\mathbf k-\mathbf k_1}
\mathcal H_{\mathbf k}
\left(z^\nodag\right)^{\mathrm T}_{\mathbf k-\mathbf k_2}.
\label{eq:supp_sb_Hk_radial}
\end{equation}

\subsection{Physical charge and spin observables}

The physical spin density is
\begin{equation}
\hat{\mathbf S}_i
=
\frac{1}{2}
\sum_{\sigma\sigma'}
c^\dagger_{i\sigma}
\boldsymbol{\tau}_{\sigma\sigma'}
c^\nodag_{i\sigma'}.
\label{eq:supp_spin_definition}
\end{equation}
Using the SRIKR constraints, it can be represented purely in terms of the
spin slave bosons as
\begin{equation}
\hat{\mathbf S}_i
=
\frac{1}{2}
\left(
p^\dagger_{0,i}\check{\mathbf p}_i
+
\check{\mathbf p}_i^\dagger p^\nodag_{0,i}
-
i\,\check{\mathbf p}_i^\dagger\times\check{\mathbf p}_i
\right),
\qquad
\check{\mathbf p}_i=(p_{1,i},-p_{2,i},p_{3,i})^{\mathrm T}.
\label{eq:supp_sb_spinop}
\end{equation}

\begin{figure}[t]
    \centering
    \includegraphics[width=\linewidth]{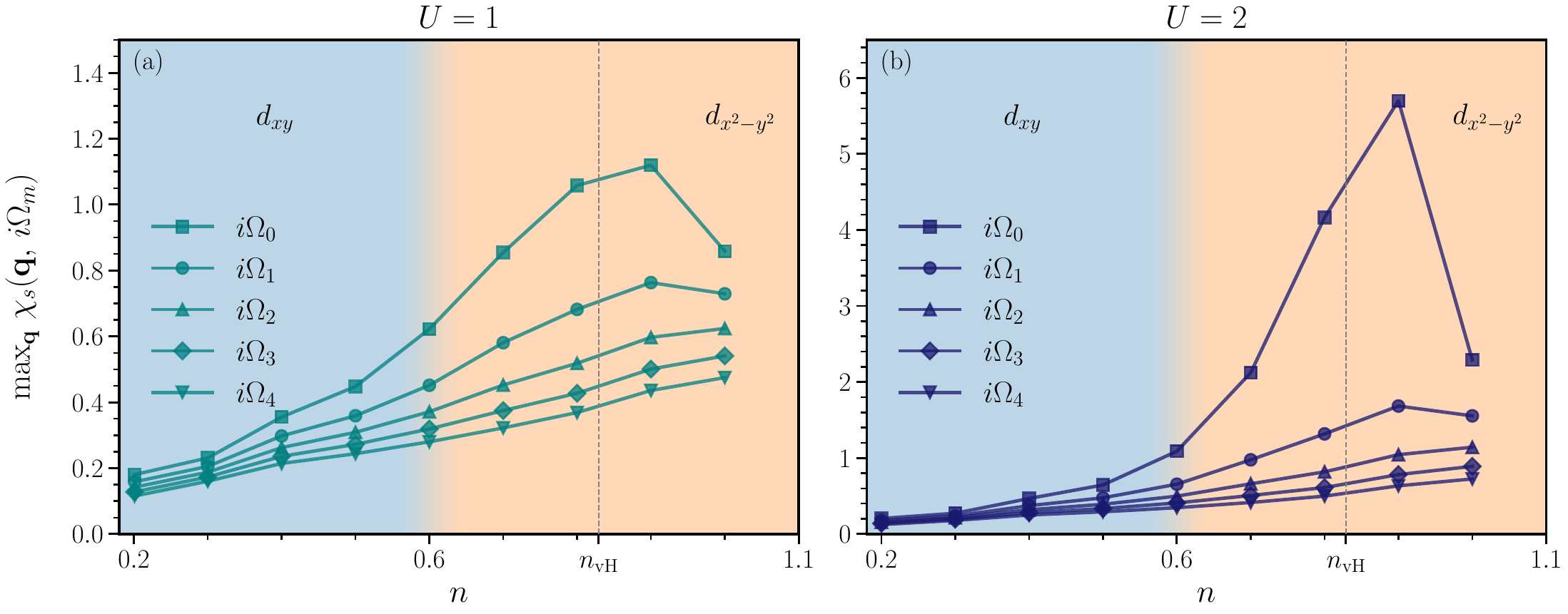}
    \caption{
    Maximum spin susceptibility
    $\max_{\mathbf q}\chi_s(\mathbf q,i\Omega_m)$ as a function of
    filling $n$ for the first five bosonic Matsubara frequencies,
    $m=0,\ldots,4$, at $T=0.02t$, for (a) $U=1$ and (b) $U=2$.
    The zero-frequency component, $i\Omega_0=0$, consistently carries
    the dominant weight. The separation from the finite-frequency
    components becomes particularly pronounced at higher fillings and
    for stronger interaction, while the nonzero-frequency components
    remain finite and contribute to the fully dynamical pairing kernel.
    Background shading indicates the leading superconducting regimes, and the vertical dashed line marks the van-Hove
    filling $n_{\mathrm{vH}}$.
    }
    \label{fig:supp_bosonic}
\end{figure}

In the radial gauge this reduces to the simple form
\begin{equation}
\hat{\mathbf S}_i \rightarrow p_{0,i}\check{\mathbf p}_i .
\label{eq:supp_spin_radial}
\end{equation}

The physical density operator is
\begin{equation}
\hat n_i
=
\sum_{\sigma}c^\dagger_{i\sigma}c^\nodag_{i\sigma}
=
\sum_{\sigma}f^\dagger_{i\sigma}f^\nodag_{i\sigma}
=
1+d_i^\dagger d_i^\nodag-e_i^\dagger e_i^\nodag .
\label{eq:supp_sb_densityop}
\end{equation}
Thus, charge fluctuations are encoded in the scalar slave-boson sector, whereas
spin fluctuations are encoded in the vector \(p_a\) sector. Expanding Eq.~\eqref{eq:supp_effective_action} around the saddle point,
\(\psi=\bar\psi+\delta\psi\), gives the Gaussian fluctuation action
\begin{equation}
\delta\mathcal S^{(2)}
=
\sum_q
\delta\psi_\mu(-q)\,
\mathcal M_{\mu\nu}(q)\,
\delta\psi_\nu(q),
\label{eq:supp_sb_quad}
\end{equation}
where \(\mathcal M(q)\) is the inverse propagator of the slave-boson
fluctuations. The physical response functions are obtained by projecting the
inverse fluctuation matrix onto the corresponding charge or spin vertices:
\begin{equation}
\chi_{s/c}(q)
=
\sum_{\mu,\nu}
\mathscr C^{\mu\nu}_{s/c}\,
\mathcal M^{-1}_{\mu\nu}(q).
\label{eq:supp_sb_chi}
\end{equation}
The coefficients \(\mathscr C^{\mu\nu}_{s/c}\) are fixed by
Eqs.~\eqref{eq:supp_sb_spinop} and \eqref{eq:supp_sb_densityop}. In the
paramagnetic saddle point, the spin and charge sectors decouple. Consequently, the transverse spin response is obtained from the
$(p_a,\beta_a)$ block of $\mathcal M^{-1}$, while the charge response
follows from the block spanned by the scalar bosons and their associated
Lagrange multipliers. Further details of this construction can be found
in Refs.~\cite{Riegler2020s,Seufert2021b}. The frequency dependence of the resulting spin response is illustrated
in Fig.~\ref{fig:supp_bosonic}, where we show
$\max_{\mathbf q}\chi_s(\mathbf q,i\Omega_m)$ for the first five bosonic
Matsubara frequencies. The zero-frequency component consistently carries
the largest weight, particularly toward higher filling and for stronger
interaction. The finite-frequency components nevertheless remain
appreciable, especially at lower filling, and contribute to the
off-diagonal coupling between different fermionic Matsubara frequencies
in the fully dynamical gap equation.

\section{Antisymmetrized pairing vertex and singlet/triplet projection}
\label{sec:supp_vertex_decoupling}

This section provides the derivation of the singlet and triplet pairing vertices used in the main text,
Eqs.~\eqref{eq:Vt_main} and \eqref{eq:Vs_main}, starting from a general antisymmetrized two-particle vertex.

\subsection{Cooper-pair vertex and fermionic antisymmetry}

We consider the effective scattering of a Cooper pair
$(\mathbf{k},\alpha;\,-\mathbf{k},\beta)\to(\mathbf{k}',\gamma;\,-\mathbf{k}',\delta)$
and define the corresponding pairing vertex
$\mathcal{V}^{\mathrm{pair}}_{\alpha\beta\gamma\delta}(k,k')$, where
$k=(i\omega_n,\mathbf{k})$ and $k'=(i\omega_{n'},\mathbf{k}')$.
Two bosonic momentum-frequency transfers appear naturally,
\[
q \equiv k-k',
\qquad
q' \equiv k+k'.
\]
Fermionic antisymmetry requires
\begin{equation}
\mathcal{V}^{\mathrm{pair}}_{\alpha\beta\gamma\delta}(k,k')
=
-\,\mathcal{V}^{\mathrm{pair}}_{\alpha\beta\delta\gamma}(k,-k').
\label{eq:supp_antisym}
\end{equation}
Since the effective interactions constructed from charge and spin susceptibilities depend only on the transferred
four-momentum, exchanging the outgoing fermions maps $q\leftrightarrow q'$.

\subsection{Density and spin decomposition}

Assuming SU(2) spin-rotation invariance, the most general direct particle-particle interaction can be written as
\[
\Gamma_{\alpha\beta\gamma\delta}(q)
=
\mathcal{V}^{nn}(q)\,\delta_{\alpha\gamma}\delta_{\beta\delta}
+
\mathcal{V}^{ss}(q)\,\boldsymbol{\tau}_{\alpha\gamma}\!\cdot\!\boldsymbol{\tau}_{\beta\delta}.
\]
Antisymmetrizing with respect to the outgoing legs yields
\begin{equation}
\begin{aligned}
\mathcal{V}^{\mathrm{pair}}_{\alpha\beta\gamma\delta}(k,k')
&=
\Gamma_{\alpha\beta\gamma\delta}(q)-\Gamma_{\alpha\beta\delta\gamma}(q') \\
&=
\Bigl[\mathcal{V}^{nn}(q)\,\delta_{\alpha\gamma}\delta_{\beta\delta}
      -\mathcal{V}^{nn}(q')\,\delta_{\alpha\delta}\delta_{\beta\gamma}\Bigr] \\
&\quad+
\Bigl[\mathcal{V}^{ss}(q)\,\boldsymbol{\tau}_{\alpha\gamma}\!\cdot\!\boldsymbol{\tau}_{\beta\delta}
      -\mathcal{V}^{ss}(q')\,\boldsymbol{\tau}_{\alpha\delta}\!\cdot\!\boldsymbol{\tau}_{\beta\gamma}\Bigr].
\end{aligned}
\label{eq:supp_Vpair_nnss}
\end{equation}

\begin{figure}[t]
    \centering
    \includegraphics[width=0.65\linewidth]{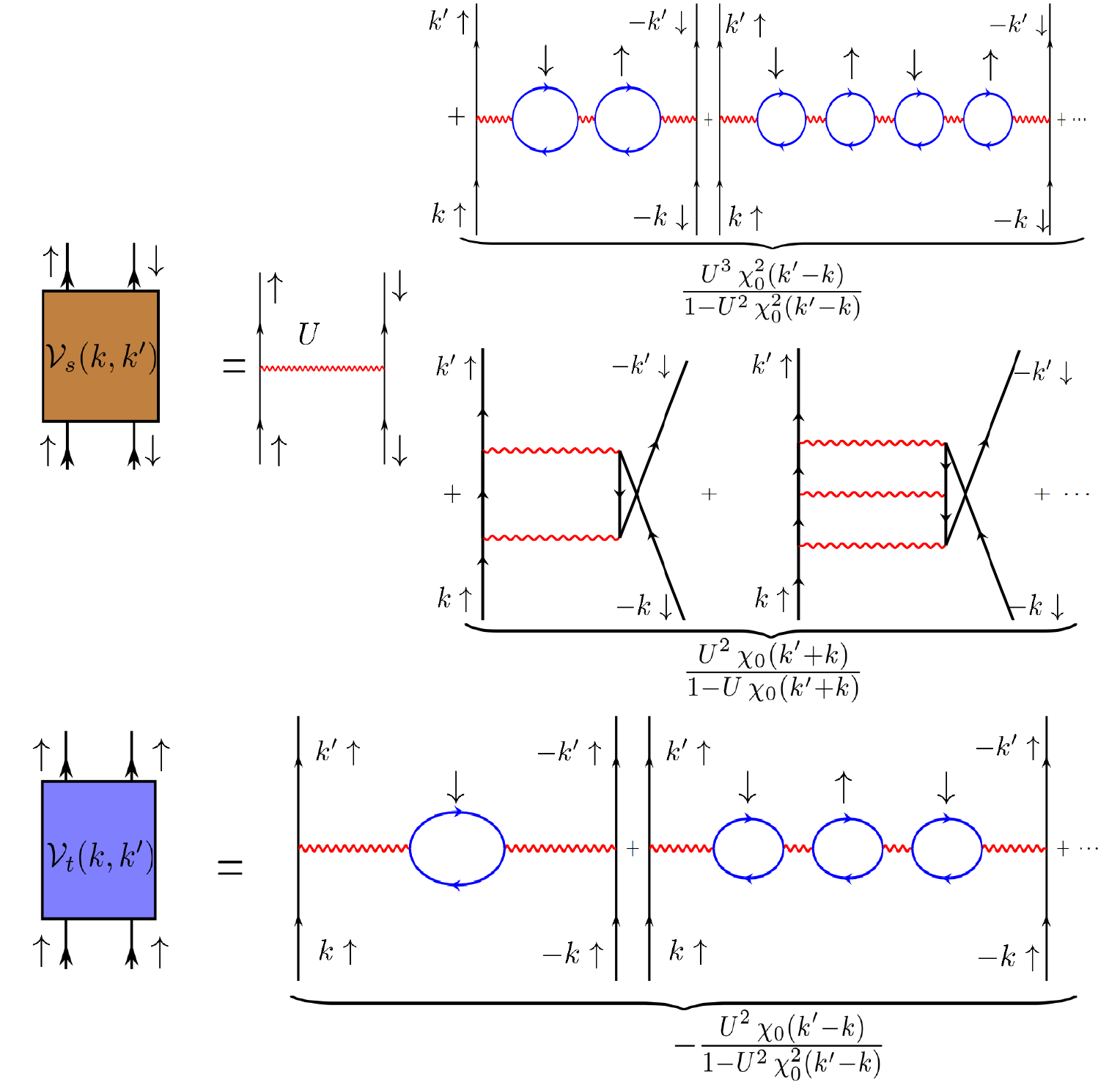}
    \caption{\textbf{Structure of the fluctuation-mediated pairing vertex.}
    Schematic diagram of the direct ($q=k-k'$) and exchange ($q'=k+k'$) contributions entering the
    antisymmetrized Cooper-pair interaction. The decomposition into density and spin channels, followed by
    projection onto singlet and triplet sectors, leads to the effective vertices in
    Eqs.~\eqref{eq:supp_Vt_chi} and \eqref{eq:supp_Vs_chi}.}
    \label{fig:supp_vertex_structure}
\end{figure}

\begin{figure}[t]
    \centering
    \includegraphics[width=0.38\columnwidth]{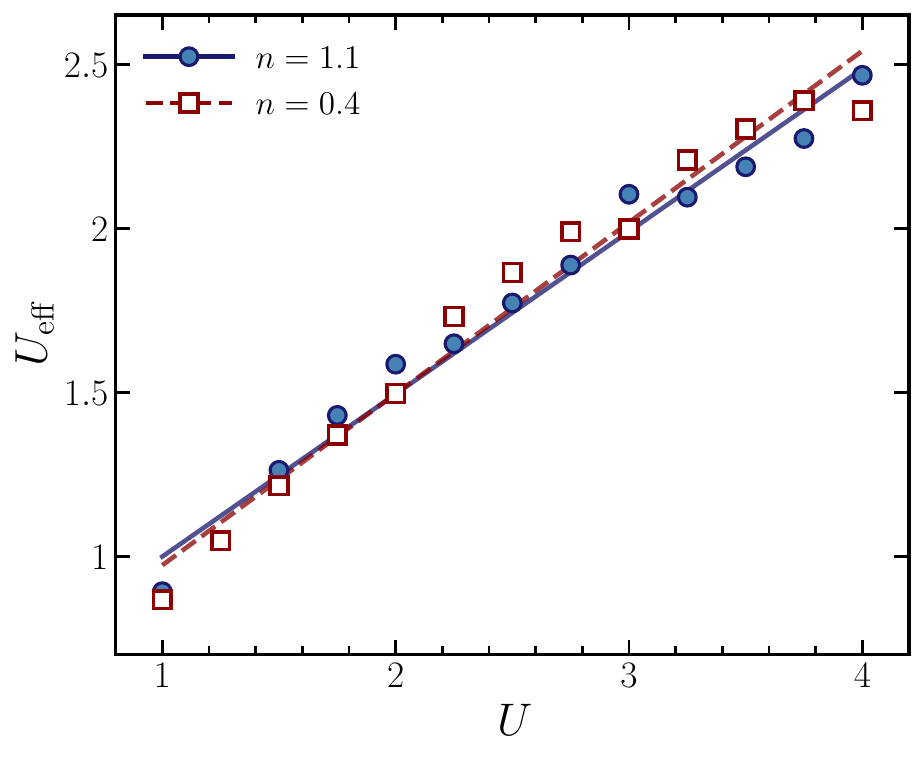}
    \caption{
    Effective local interaction $U_{\mathrm{eff}}$ as a function of the
    bare Hubbard interaction $U$ for fillings $n=0.4$ and $n=1.1$.
    Symbols denote the values extracted from
    $\mathcal{K}_s(\mathbf Q,0)$, while the lines are linear fits.
    For both fillings, $U_{\mathrm{eff}}$ exhibits an approximately
    linear dependence on $U$ over the displayed range.
    }
    \label{fig:Ueff_U}
\end{figure}

\subsection{Singlet/triplet projection}

Define the standard singlet and triplet pairing matrices
\[
\Phi^{(s)}_{\alpha\beta}\equiv (i\tau^y)_{\alpha\beta},
\qquad
\Phi^{(t),i}_{\alpha\beta}\equiv (i\tau^y\tau^i)_{\alpha\beta},
\qquad i=x,y,z.
\]
The SU(2)-invariant decomposition of the Cooper vertex then reads
\begin{equation}
\mathcal{V}^{\mathrm{pair}}_{\alpha\beta\gamma\delta}(k,k')
=
\mathcal{V}_s(k,k')\,\Phi^{(s)}_{\alpha\beta}\Phi^{(s)\dagger}_{\gamma\delta}
+
\mathcal{V}_t(k,k')\sum_{i=x,y,z}\Phi^{(t),i}_{\alpha\beta}\Phi^{(t),i\dagger}_{\gamma\delta}.
\label{eq:supp_Vpair_st}
\end{equation}

Projecting Eq.~\eqref{eq:supp_Vpair_nnss} onto equal-spin and opposite-spin channels gives
\begin{equation}
\mathcal{V}_t(k,k')
=
\frac{1}{2}\bigl[\mathcal{V}^{nn}(q)-\mathcal{V}^{nn}(q')\bigr]
+
\frac{1}{2}\bigl[\mathcal{V}^{ss}(q)-\mathcal{V}^{ss}(q')\bigr],
\label{eq:supp_Vt}
\end{equation}
and
\begin{equation}
\mathcal{V}_s(k,k')
=
\frac{1}{2}\bigl[\mathcal{V}^{nn}(q)+\mathcal{V}^{nn}(q')\bigr]
-\frac{3}{2}\bigl[\mathcal{V}^{ss}(q)+\mathcal{V}^{ss}(q')\bigr].
\label{eq:supp_Vs}
\end{equation}
The exchange terms proportional to $q'$ implement the antisymmetry constraint
of Eq.~\eqref{eq:supp_antisym}.

\subsection{Second-order fluctuation-mediated approximation}

Within the spin-fluctuation framework used in the main text, we
approximate the effective interaction to second order in the spin- and
charge-channel kernels $\mathcal{K}_{s,c}(q)$. Writing again
$q=k-k'$ and $q'=k+k'$, the triplet vertex becomes
\begin{align}
\mathcal{V}_t(k,k')
=
\frac{1}{4}\Bigl[
&\mathcal{K}_s^2(q')\chi_s(q')
+\mathcal{K}_c^2(q')\chi_c(q')
\nonumber\\
&-\mathcal{K}_s^2(q)\chi_s(q)
-\mathcal{K}_c^2(q)\chi_c(q)
\Bigr],
\label{eq:supp_Vt_chi}
\end{align}
while the singlet vertex is
\begin{align}
\mathcal{V}_s(k,k')
=
\mathcal{K}_{\mathrm{irr}}(q)
&-\frac{1}{4}\Bigl[
\mathcal{K}_c^2(q)\chi_c(q)
+\mathcal{K}_c^2(q')\chi_c(q')
\Bigr]
\nonumber\\
&+\frac{3}{4}\Bigl[
\mathcal{K}_s^2(q)\chi_s(q)
+\mathcal{K}_s^2(q')\chi_s(q')
\Bigr].
\label{eq:supp_Vs_chi}
\end{align}
For a local Hubbard interaction, the irreducible contribution
$\mathcal{K}_{\mathrm{irr}}(q)$ enters only the singlet channel, while
the $q'$ terms implement the antisymmetrization required by
Eq.~\eqref{eq:supp_antisym}. Our choice of vertex structure is
motivated by the effective RPA summation of bubble and ladder diagrams
shown schematically in Fig.~\ref{fig:supp_vertex_structure}; expanding
the corresponding RPA-dressed interactions to second order in the
channel kernels yields Eqs.~\eqref{eq:supp_Vt_chi} and
\eqref{eq:supp_Vs_chi}.
For the numerical solution, we employ a local-vertex approximation and
extract an effective onsite scale from the spin-channel kernel,
\begin{equation}
U_{\mathrm{eff}}(n,U,T)
=
\mathcal{K}_s(\mathbf{Q},0),
\qquad
\mathbf{Q}
=
\underset{\mathbf q}{\operatorname{arg\,max}}\,
\chi_s(\mathbf q,0),
\label{eq:supp_Ueff_definition}
\end{equation}
where
\begin{equation}
\mathcal{K}_s(q)
=
\chi_0^{-1}(q)-\chi_s^{-1}(q)
=
\frac{\chi_s(q)-\chi_0(q)}
{\chi_s(q)\chi_0(q)}.
\label{eq:supp_Ks_definition}
\end{equation}
The resulting $U_{\mathrm{eff}}$ is therefore obtained from the
slave-boson response functions and is recalculated for every filling
and temperature entering the $T_c$ sweep. Fig.~\ref{fig:Ueff_U} depicts $U_{\mathrm{eff}}$ scales approximately linearly with $U$ for general fillings.

\section{Mean-field equations}
\label{sec:supp_mf_eqs}

This section summarizes the static mean-field gap equation that follows from the pairing interaction discussed above
and that is referred to in the main text when discussing the static approximation to the full dynamic gap equation.

\subsection{BCS mean-field decoupling}

We start from a generic momentum-space pairing Hamiltonian,
\begin{equation}
H = \sum_{\mathbf{k},\sigma}\xi(\mathbf{k})\,c^\dagger_{\mathbf{k}\sigma}c_{\mathbf{k}\sigma}
+\frac{1}{2}\sum_{\mathbf{k}\mathbf{k}'}\sum_{\alpha\beta\gamma\delta}
\mathcal{V}_{\alpha\beta\gamma\delta}(\mathbf{k},\mathbf{k}')
\,c^\dagger_{\mathbf{k}\alpha}c^\dagger_{-\mathbf{k}\beta}
c_{-\mathbf{k}'\delta}c_{\mathbf{k}'\gamma},
\label{eq:supp_H_pairing_general}
\end{equation}
where $\xi(\mathbf{k})=\varepsilon(\mathbf{k})-\mu$.
The superconducting order parameter is defined as
\begin{equation}
\Delta_{\alpha\beta}(\mathbf{k}) \equiv
-\sum_{\mathbf{k}'\gamma\delta}\mathcal{V}_{\alpha\beta\gamma\delta}(\mathbf{k},\mathbf{k}')
\,\langle c_{-\mathbf{k}'\delta}\,c_{\mathbf{k}'\gamma}\rangle.
\label{eq:supp_gap_def}
\end{equation}
Performing the standard BCS decoupling gives the mean-field Hamiltonian
\begin{equation}
\begin{aligned}
H_{\mathrm{MF}}
&=\sum_{\mathbf{k},\sigma}
\xi(\mathbf{k})\,
c^\dagger_{\mathbf{k}\sigma}c_{\mathbf{k}\sigma}
-\frac{1}{2}\sum_{\mathbf{k}}\sum_{\alpha\beta}
\Bigl[
c^\dagger_{\mathbf{k}\alpha}
c^\dagger_{-\mathbf{k}\beta}\,
\Delta_{\alpha\beta}(\mathbf{k})
+
\Delta^*_{\alpha\beta}(\mathbf{k})\,
c_{-\mathbf{k}\beta}c_{\mathbf{k}\alpha}
\Bigr]
\\
&\quad
-\frac{1}{2}
\sum_{\mathbf{k}\mathbf{k}'}
\sum_{\alpha\beta\gamma\delta}
\Delta^*_{\alpha\beta}(\mathbf{k})
\bigl[\mathcal{V}^{-1}\bigr]_{\alpha\beta\gamma\delta}
(\mathbf{k},\mathbf{k}')
\Delta_{\gamma\delta}(\mathbf{k}') .
\end{aligned}
\label{eq:supp_H_MF}
\end{equation}
For pure singlet or triplet states the quasiparticle energy reduces to
\[
E_{s/t}(\mathbf{k})=\sqrt{\xi(\mathbf{k})^2+|\Delta_{s/t}(\mathbf{k})|^2}.
\]
\subsection{Static finite-temperature gap equation}
\label{static gap}

Using Eq.~\eqref{eq:supp_gap_def} together with the anomalous thermal expectation value, the finite-temperature
self-consistency condition can be written as
\begin{equation}
\Delta_{\alpha\beta}(\mathbf{k})
=
-\sum_{\mathbf{k}'\gamma\delta}
\mathcal{V}_{\alpha\beta\gamma\delta}(\mathbf{k},\mathbf{k}')
\frac{\Delta_{\gamma\delta}(\mathbf{k}')}{2E(\mathbf{k}')}
\tanh\!\left(\frac{E(\mathbf{k}')}{2T}\right).
\label{eq:supp_gap_general}
\end{equation}
Projecting the gap into singlet and triplet sectors,
\begin{equation}
\Delta_{\alpha\beta}(\mathbf{k})
=
\Delta_s(\mathbf{k})\,(i\tau^y)_{\alpha\beta}
+
\Delta_t^i(\mathbf{k})\,(i\tau^y\tau^i)_{\alpha\beta},
\label{eq:supp_gap_decomp}
\end{equation}
one obtains the decoupled static gap equations
\begin{equation}
\Delta_{s/t}(\mathbf{k})
=
-\sum_{\mathbf{k}'}
\mathcal{V}_{s/t}(\mathbf{k},\mathbf{k}')
\,\frac{\Delta_{s/t}(\mathbf{k}')}{2E_{s/t}(\mathbf{k}')}
\tanh\!\left(\frac{E_{s/t}(\mathbf{k}')}{2T}\right),
\label{eq:supp_gap_st}
\end{equation}
with
\begin{equation}
E_{s/t}(\mathbf{k})=\sqrt{\xi(\mathbf{k})^2+|\Delta_{s/t}(\mathbf{k})|^2}.
\label{eq:supp_quasi_energy}
\end{equation}
Equation~\eqref{eq:supp_gap_st} is the static limit of the full dynamic gap equation,
Eq.~\eqref{eq:gap_dynamic_main} in the main text, obtained by neglecting the bosonic-frequency dependence
of the interaction and the fermionic-frequency dependence of the gap.

\subsection{Parity constraints and free energy}

The Pauli principle requires
\begin{equation}
\Delta_{\alpha\beta}(\mathbf{k})=-\Delta_{\beta\alpha}(-\mathbf{k}).
\label{eq:supp_pauli}
\end{equation}
Therefore a pure singlet state satisfies
\[
\Delta_s(\mathbf{k})=\Delta_s(-\mathbf{k}),
\]
while a pure triplet state satisfies
\[
\Delta_t^i(\mathbf{k})=-\Delta_t^i(-\mathbf{k}).
\]

The corresponding mean-field free energy at the saddle point is
\begin{equation}
\begin{aligned}
\mathcal{F}
&=\sum_{\mathbf{q}}
\Biggl\{
-2T\,\ln\!\bigl[1+e^{-\beta E_{s/t}(\mathbf{q})}\bigr]
+\frac{1}{2}\,\frac{|\Delta_{s/t}(\mathbf{q})|^2}{E_{s/t}(\mathbf{q})}
\tanh\!\Bigl(\frac{E_{s/t}(\mathbf{q})}{2T}\Bigr)
- E_{s/t}(\mathbf{q})+\xi(\mathbf{q})
\Biggr\}
+\mu\,N,
\end{aligned}
\label{eq:supp_free_energy}
\end{equation}

\section{Expansion in the basis and numerical solution}
\label{sec:supp_basis_expansion}

To solve the full dynamic gap equation, Eq.~\eqref{eq:gap_dynamic_main}
in the main text, we expand the gap function in a set of lattice harmonics
and determine the corresponding Matsubara-frequency-dependent coefficients
self-consistently.

\subsection{Form-factor basis}
\label{sec:form_factor_basis}

We expand the gap function in a finite set of lattice harmonics,
\begin{equation}
\Delta(\mathbf{k},i\omega_n)
=
\sum_{\alpha=1}^{M}
c_\alpha(i\omega_n)\,\varphi_\alpha(\mathbf{k}),
\label{eq:supp_gap_basis_expansion}
\end{equation}
where \(c_\alpha(i\omega_n)\) are frequency-dependent expansion
coefficients and \(\varphi_\alpha(\mathbf{k})\) denotes the corresponding
momentum-space form factor. The form-factor basis used in this work is
summarized in Table~\ref{tab:form_factor_basis}.
\begin{table}[H]
\centering
\caption{
Form-factor basis used in the expansion of
\(\Delta(\mathbf{k},i\omega_n)\). The irreducible representations are
given for the square-lattice point group with inversion parity.
}
\label{tab:form_factor_basis}
\scriptsize
\renewcommand{\arraystretch}{1.5}
\setlength{\tabcolsep}{5pt}
\resizebox{0.56\columnwidth}{!}{%
\begin{tabular}{c|c|c|c}
\noalign{\hrule height 1.2pt}

Channel \(\alpha\)
&
Form factor \(\varphi_\alpha(\mathbf{k})\)
&
Parity
&
Irrep
\\
\hline
\rowcolor{Gray}
\(s\)
&
\(\displaystyle 1\)
&
even
&
\(A_{1g}\)
\\

\(s^{(1)}\)
&
\(\displaystyle \cos k_x+\cos k_y\)
&
even
&
\(A_{1g}\)
\\

\rowcolor{Gray}
\(s^{(2)}\)
&
\(\displaystyle \cos 2k_x+\cos 2k_y\)
&
even
&
\(A_{1g}\)
\\

\(d_{x^2-y^2}^{(1)}\)
&
\(\displaystyle \cos k_x-\cos k_y\)
&
even
&
\(B_{1g}\)
\\

\rowcolor{Gray}
\(d_{xy}^{(1)}\)
&
\(\displaystyle \sin k_x\sin k_y\)
&
even
&
\(B_{2g}\)
\\

\(d_{xy}^{(2)}\)
&
\(\displaystyle
\sin k_x\sin 2k_y
+
\sin 2k_x\sin k_y
\)
&
even
&
\(B_{2g}\)
\\

\rowcolor{Gray}
\(d_{x^2-y^2}^{(2)}\)
&
\(\displaystyle \cos 2k_x-\cos 2k_y\)
&
even
&
\(B_{1g}\)
\\

\(d_{x^2-y^2}^{(3)}\)
&
\(\displaystyle
\cos 2k_x\cos k_y
-
\cos 2k_y\cos k_x
\)
&
even
&
\(B_{1g}\)
\\

\rowcolor{Gray}
\(g\)
&
\(\displaystyle
(\cos k_x-\cos k_y)\sin k_x\sin k_y
\)
&
even
&
\(A_{2g}\)
\\

\(p_x\)
&
\(\displaystyle \sin k_x\)
&
odd
&
\(E_u\)
\\

\rowcolor{Gray}
\(p_y\)
&
\(\displaystyle \sin k_y\)
&
odd
&
\(E_u\)
\\

\(p'_x\)
&
\(\displaystyle
(\cos k_x-\cos k_y)\sin k_x
\)
&
odd
&
\(E_u\)
\\

\rowcolor{Gray}
\(p'_y\)
&
\(\displaystyle
-(\cos k_x-\cos k_y)\sin k_y
\)
&
odd
&
\(E_u\)
\\

\(\vdots\)
&
\(\vdots\)
&
\(\vdots\)
&
\(\vdots\)
\\

\noalign{\hrule height 1.2pt}
\end{tabular}%
}
\end{table}

The basis contains onsite and extended \(s\)-wave harmonics, several
\(d\)-wave harmonics, a \(g\)-wave harmonic, and odd-parity \(p\)-wave
form factors. The extended \(p'\)-wave sector is included as the doublet
\((p'_x,~p'_y)\), so that the basis remains closed under the square-lattice
point-group operations.

\subsection{Projection of the dynamic gap equation}
\label{sec:supp_basis_projection}

Projecting Eq.~\eqref{eq:gap_dynamic_main} onto the basis functions gives
\begin{equation}
c_\beta(i\omega_n)
=
-\frac{T}{\mathcal{N}_\beta}
\sum_{m,\alpha}
\Gamma_{\beta\alpha}(n,m;\{c\})\,c_\alpha(i\omega_m),
\label{eq:supp_nonlin_proj}
\end{equation}
where
\[
\mathcal{N}_\beta
=
\int_{\mathrm{BZ}}\frac{d^2k}{(2\pi)^2}\,|\varphi_\beta(\mathbf{k})|^2
\]
and
\begin{equation}
\Gamma_{\beta\alpha}(n,m;\{c\})
=
\int_{\mathrm{BZ}}\frac{d^2k\,d^2p}{(2\pi)^4}\,
\varphi_\beta^*(\mathbf{k})\,
\mathcal{V}(\mathbf{k}-\mathbf{p},i\omega_n-i\omega_m)\,
\varphi_\alpha(\mathbf{p})\,
\frac{1}{\omega_m^2+\xi_{\mathbf p}^2+\left|\sum_\gamma c_\gamma(i\omega_m)\varphi_\gamma(\mathbf p)\right|^2}.
\label{eq:supp_Gamma_def}
\end{equation}
Equations~\eqref{eq:supp_nonlin_proj} and \eqref{eq:supp_Gamma_def} define the nonlinear fixed-point problem for
the coefficient vector $\mathbf c(i\omega_n)$.

\subsection{Iterative fixed-point solver}
\label{sec:supp_fixed_point_solver}

We solve Eq.~\eqref{eq:supp_nonlin_proj} iteratively using linear mixing. Starting from an initial seed
$\mathbf c^{(0)}(i\omega_n)$, the updated coefficients are obtained from the fixed-point map defined by
Eq.~\eqref{eq:supp_nonlin_proj}, followed by a mixing step
\[
c_\beta^{(t)}(i\omega_n)
=
(1-\eta)c_\beta^{(t-1)}(i\omega_n)
+
\eta\,\widetilde c_\beta^{(t)}(i\omega_n),
\]
with mixing parameter $\eta\in(0,1]$. The iteration is stopped once the Euclidean norm of the difference between
successive iterates falls below a prescribed tolerance.

\begin{enumerate}
    \item Initialize the coefficient vector $\mathbf c^{(0)}(i\omega_n)$.

    \item For $t=1,2,\dots,\mathbb{T}_{\max}$, construct
    \[
    \Delta^{(t-1)}(\mathbf p,i\omega_m)
    =
    \sum_\gamma c_\gamma^{(t-1)}(i\omega_m)\,\varphi_\gamma(\mathbf p).
    \]

    \item Using $\Delta^{(t-1)}(\mathbf p,i\omega_m)$, assemble the kernel
    \[
    \Gamma_{\beta\alpha}(n,m;\{\mathbf c^{(t-1)}\})
    \]
    from Eq.~\eqref{eq:supp_Gamma_def}.

    \item Compute the updated coefficients $\widetilde c_\beta^{(t)}(i\omega_n)$ from Eq.~\eqref{eq:supp_nonlin_proj}.

    \item Apply linear mixing according to
    \[
    c_\beta^{(t)}
    =
    (1-\eta)c_\beta^{(t-1)}+\eta \widetilde c_\beta^{(t)}.
    \]

    \item Check for convergence. If
    \[
    \|\mathbf c^{(t)}-\mathbf c^{(t-1)}\|_2 < \mathrm{tol},
    \]
    then stop and take $\mathbf c^{(t)}$ as the self-consistent solution.

    \item If convergence is not reached, continue the iteration until $t=\mathbb{T}_{\max}$.

    \item If the procedure does not converge within $\mathbb{T}_{\max}$ steps, take
    \[
    \mathbf c^{(\mathbb{T}_{\max})}
    \]
    as the final approximate solution.
\end{enumerate}

\subsection{Identification of the leading symmetry channel}

Once a converged solution is obtained, we identify the leading pairing symmetry from the largest overlap with the
basis functions at the lowest Matsubara frequency,
\begin{equation}
\alpha_{\mathrm{lead}}
=
\arg\max_\alpha \left|\mathrm{overlap}_\alpha(i\omega_0)\right|,
\qquad
\left|\mathrm{overlap}_{\mathrm{lead}}\right|
=
\max_\alpha \left|\mathrm{overlap}_\alpha(i\omega_0)\right|.
\label{eq:supp_lead_overlap}
\end{equation}

\begin{figure}[t]
  \centering
  \begin{minipage}[b]{0.48\linewidth}
    \centering
    \includegraphics[width=\linewidth]{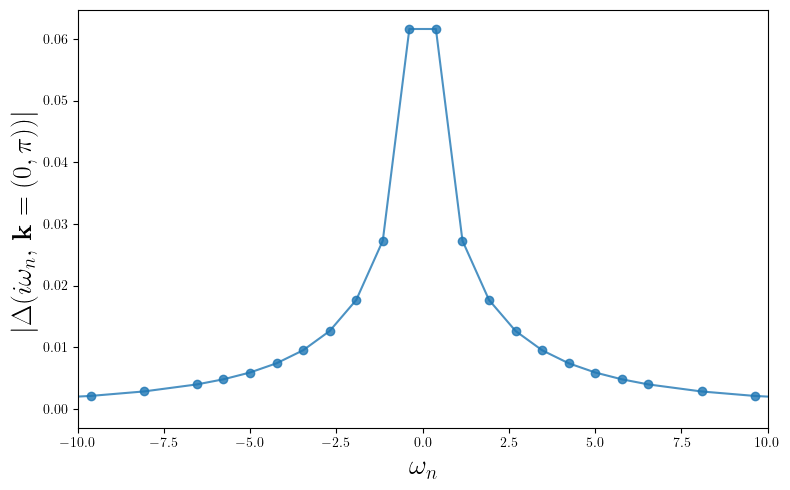}
    \caption{Representative converged Matsubara-frequency dependence of the leading superconducting gap amplitude at $\mathbf{k}=(0,\pi)$.}
    \label{fig:supp_gap_freq}
  \end{minipage}
  \hfill
  \begin{minipage}[b]{0.48\linewidth}
    \centering
    \includegraphics[width=0.75\linewidth,height=0.62\linewidth]{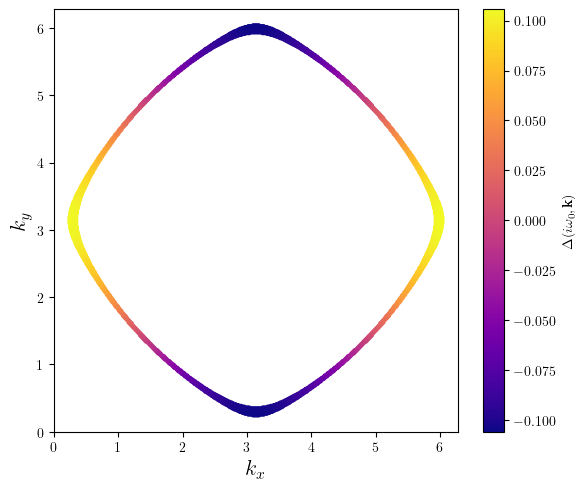}
    \caption{Projection of $d_{x^2-y^2}$ gap structure onto the Fermi surface.}
    \label{fig:supp_gap_fs}
  \end{minipage}
\end{figure}

Figures~\ref{fig:supp_gap_freq} and \ref{fig:supp_gap_fs} show a representative converged solution and the
associated Fermi-surface gap structure.

\begin{figure}[t]
\centering
\includegraphics[page=1,width=0.72\linewidth]{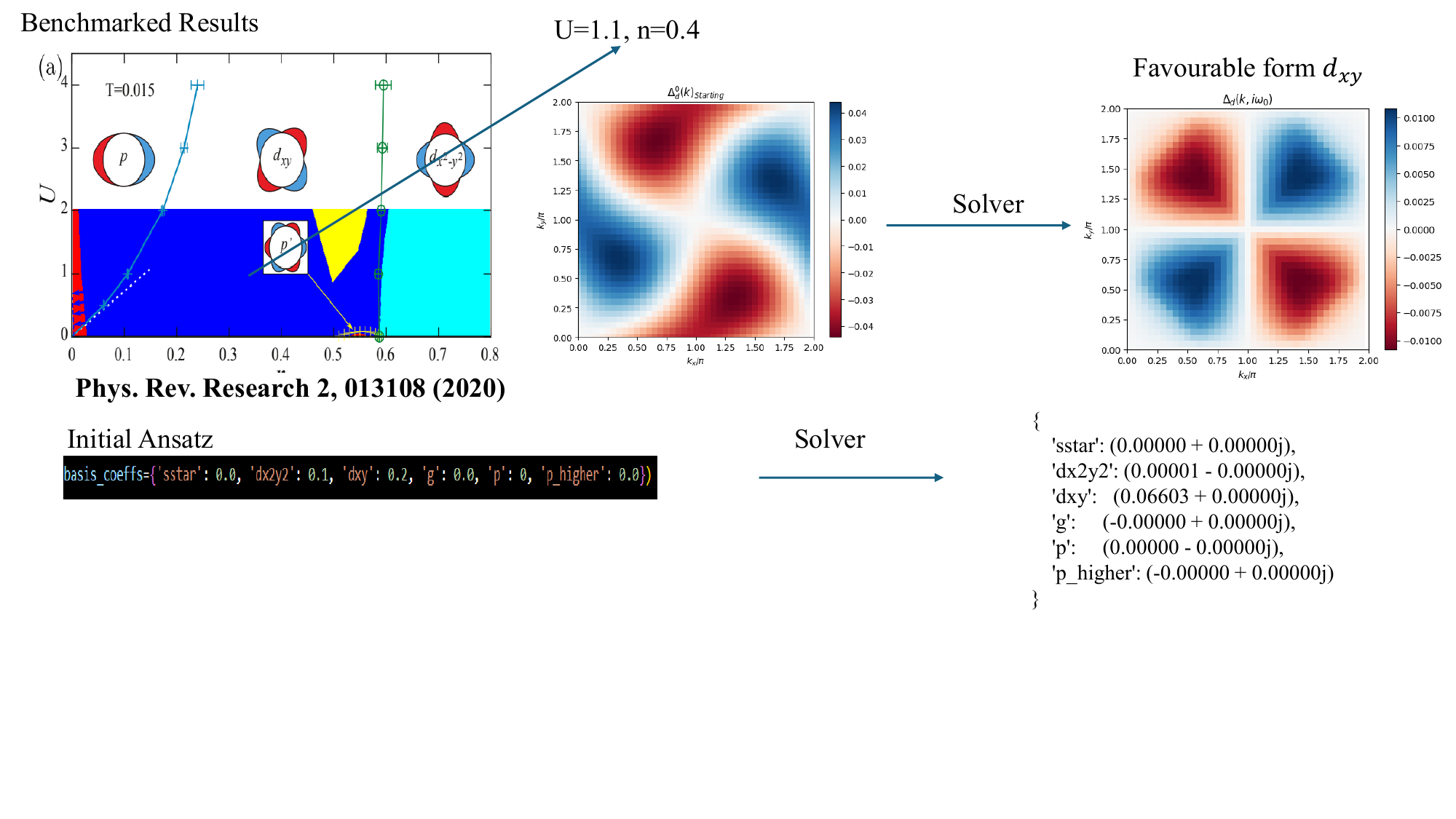}

\vspace{0.1em}

\includegraphics[page=2,width=0.72\linewidth]{Figures/page_1.pdf}
\caption{\textbf{Benchmark of the solver against the FLEX+RPA phase diagram.}
The phase diagram shown in the left panels is taken from Ref.~\cite{Romer2020p}. 
For the filling $n=0.4$ \& $n=0.7$ and interaction strength $U=1.1$, two different initial ans\"atze were supplied to the solver in order to test whether the expected pairing symmetry is recovered. 
\textbf{Top row:} starting from an initial ansatz with dominant $d_{xy}$ component, the solver converges to a gap function with the characteristic $d_{xy}$ sign structure, consistent with the favourable pairing channel in that region of the phase diagram. 
\textbf{Bottom row:} starting from an initial ansatz with dominant $d_{x^{2}-y^{2}}$ component, the solver correspondingly converges to a gap function with $d_{x^{2}-y^{2}}$ symmetry. 
The rightmost panels show the converged momentum dependence of the gap function $\Delta(\mathbf{k})$, while the printed coefficient sets display the numerical weights of the symmetry components after convergence. This comparison serves as a benchmark that the solver correctly identifies and stabilizes the intended pairing representation when initialized in the corresponding channel.}
\label{fig:flex_bench_combined}
\end{figure}
To benchmark the qualitative frequency and symmetry structure obtained from the slave-boson-based pairing kernel,
we also implemented a standard single-band FLEX+RPA calculation.

\subsection{Working equations}

We use the square-lattice dispersion
\begin{equation}
\varepsilon_{\mathbf{k}}
=
-2t\big(\cos k_x+\cos k_y\big)-4t'\cos k_x\cos k_y,
\label{eq:flex_dispersion}
\end{equation}
and define $\xi_{\mathbf k}=\varepsilon_{\mathbf k}-\mu$. With
$k=(\mathbf{k},i\omega_n)$ and $q=(\mathbf{q},i\nu_m)$, the dressed Green's function is
\begin{equation}
G(k)=\frac{1}{i\omega_n-\xi_{\mathbf{k}}-\Sigma(k)}.
\label{eq:flex_dyson}
\end{equation}
The irreducible bubble is
\begin{equation}
\chi^{0}(q)
=
-\frac{T}{N}\sum_{k}G(k+q)\,G(k),
\label{eq:flex_chi0}
\end{equation}
and the RPA-resummed spin and charge susceptibilities are
\begin{equation}
\chi^{s}(q)=\frac{\chi^{0}(q)}{1-U\chi^{0}(q)},
\qquad
\chi^{c}(q)=\frac{\chi^{0}(q)}{1+U\chi^{0}(q)}.
\label{eq:flex_rpa_chi}
\end{equation}
The FLEX effective interaction and the self-energy update are
\begin{equation}
\mathcal{V}_{\mathrm{eff}}(q)
=
\frac{3}{2}U^{2}\chi^{s}(q)
+\frac{1}{2}U^{2}\chi^{c}(q)
-U^{2}\chi^{0}(q)
+U,
\label{eq:flex_veff}
\end{equation}
\begin{equation}
\Sigma(k)
=
\frac{T}{N}\sum_{q}\mathcal{V}_{\mathrm{eff}}(q)\,G(k-q).
\label{eq:flex_sigma}
\end{equation}
The singlet pairing interaction used in the linearized gap equation is
\begin{equation}
\mathcal{V}^{(s)}(q)
=
\frac{3}{2}U^{2}\chi^{s}(q)
-\frac{1}{2}U^{2}\chi^{c}(q)
+U,
\label{eq:flex_vs}
\end{equation}
and the leading eigenvalue $\lambda$ and eigenfunction $\Delta(k)$ satisfy
\begin{equation}
\lambda\,\Delta(k)
=
-\frac{T}{N}\sum_{k'}\mathcal{V}^{(s)}(k-k')\,G(k')\,G(-k')\,\Delta(k').
\label{eq:flex_gap}
\end{equation}
The superconducting transition temperature is estimated from $\lambda(T_c)=1$.

\subsection{Benchmark results}

Representative leading eigenfunctions from the FLEX+RPA calculation are shown in
Figs.~\ref{fig:flex_bench_combined}. In both cases, the extracted gap structure agrees
with the target square-lattice symmetry and serves as a benchmark for the symmetry assignment used in the main text.

\section{Computational workflow of the superconducting calculation}
\label{sec:computational_workflow}

In this section, we summarize the numerical procedure used to obtain
the superconducting phase diagram and transition temperatures presented
in the main text. The complete workflow is illustrated in
Fig.~\ref{fig:computational_workflow}.

\begin{figure*}[t]
\centering

\resizebox{0.82\textwidth}{!}{%
\begin{tikzpicture}[
    font=\scriptsize,
    node distance=4.5mm and 9mm,
    >=Latex,
    box/.style={
        rectangle,
        rounded corners=2pt,
        draw,
        thick,
        align=center,
        minimum height=8mm,
        text width=4.3cm,
        inner sep=3pt
    },
    widebox/.style={
        rectangle,
        rounded corners=2pt,
        draw,
        thick,
        align=center,
        minimum height=10mm,
        text width=6.2cm,
        inner sep=3pt
    },
    decision/.style={
        diamond,
        draw,
        thick,
        align=center,
        aspect=2.2,
        text width=2.3cm,
        inner sep=1pt
    },
    output/.style={
        rectangle,
        rounded corners=2pt,
        draw,
        double,
        thick,
        align=center,
        minimum height=8mm,
        text width=4.7cm,
        inner sep=3pt
    },
    arrow/.style={
        ->,
        thick
    },
    loop/.style={
        ->,
        thick,
        dashed
    }
]


\node[box] (input) {
\textbf{Input parameters}\\
$(n,U,T,t,t')$
};

\node[widebox, below=of input] (saddle) {
\textbf{Paramagnetic SRIKR saddle point}\\
Solve self-consistently for the slave-boson amplitudes and
constraint fields; obtain the renormalized dispersion
$\widetilde{\varepsilon}_{\mathbf k}$
};

\node[widebox, below=of saddle] (susceptibility) {
\textbf{Gaussian slave-boson fluctuations}\\
Construct $M_{\mu\nu}(q)$ and calculate
$\chi_s(q)$, $\chi_c(q)$, and the renormalized bubble $\chi_0(q)$
};

\node[widebox, below=of susceptibility] (vertex) {
\textbf{Effective local vertex ($\mathcal{V}$)}\\
Identify the dominant static magnetic wave vector $\mathbf Q$ and
evaluate
$\displaystyle
U_{\mathrm{eff}}(n,U,T)=K_s(\mathbf Q,0)
$
};


\node[box, below left=7mm and 5mm of vertex] (dynamic) {
\textbf{Dynamical kernel}\\[1mm]
$\chi_{s/c}(\mathbf q,i\Omega_m)$\\
Full momentum and bosonic-frequency dependence
};

\node[box, below right=7mm and 5mm of vertex] (static) {
\textbf{Static kernel}\\[1mm]
$\chi_{s/c}(\mathbf q,i\Omega_m)
\rightarrow
\chi_{s/c}(\mathbf q,0)$
};

\node[widebox, below=12mm of $(dynamic.south)!0.5!(static.south)$]
(gapansatz) {
\textbf{Form-factor ansatz}\\
$\displaystyle
\Delta(\mathbf k,i\omega_n)
=
\sum_{\alpha}
c_{\alpha}(i\omega_n)\phi_{\alpha}(\mathbf k)
$\\
Initialize $c_{\alpha}^{(0)}(i\omega_n)$
};


\node[widebox, below=of gapansatz] (gapsolver) {
\textbf{Anisotropic gap-equation iteration}\\
Insert $\Delta^{(r)}(\mathbf k,i\omega_n)$ into the static or
dynamical gap equation and calculate
$\Delta^{(r+1)}(\mathbf k,i\omega_n)$
};

\node[decision, below=7mm of gapsolver] (gapconverged) {
Gap converged?
};

\node[box, right=15mm of gapconverged] (gapupdate) {
Update the coefficients\\
$c_{\alpha}^{(r)}
\rightarrow
c_{\alpha}^{(r+1)}$
};

\node[output, below=8mm of gapconverged] (gapsolution) {
\textbf{Converged solution}\\
$\Delta(\mathbf k,i\omega_n)$ and leading pairing symmetry
};


\node[decision, below=8mm of gapsolution] (temperature) {
$T_c$ located?
};

\node[box, right=15mm of temperature] (nexttemperature) {
Choose the next temperature\\
$T\rightarrow T+\delta T$
};

\node[output, below=8mm of temperature] (finaloutput) {
\textbf{Final results}\\
$T_c(n,U)$, leading symmetry, and phase diagram
};


\draw[arrow] (input) -- (saddle);
\draw[arrow] (saddle) -- (susceptibility);
\draw[arrow] (susceptibility) -- (vertex);

\draw[arrow] (vertex.south) |- (dynamic.north);
\draw[arrow] (vertex.south) |- (static.north);

\draw[arrow] (dynamic.south) |- (gapansatz.north);
\draw[arrow] (static.south) |- (gapansatz.north);

\draw[arrow] (gapansatz) -- (gapsolver);
\draw[arrow] (gapsolver) -- (gapconverged);

\draw[arrow]
(gapconverged.east)
--
node[above] {no}
(gapupdate.west);

\draw[loop]
(gapupdate.north)
|- ($(gapsolver.east)+(4mm,0)$)
-- (gapsolver.east);

\draw[arrow]
(gapconverged.south)
--
node[right] {yes}
(gapsolution.north);

\draw[arrow] (gapsolution) -- (temperature);

\draw[arrow]
(temperature.east)
--
node[above] {no}
(nexttemperature.west);

\draw[loop]
(nexttemperature.east)
-- ++(10mm,0)
|- ($(saddle.east)+(5mm,0)$)
-- (saddle.east);

\draw[arrow]
(temperature.south)
--
node[right] {yes}
(finaloutput.north);

\end{tikzpicture}%
}

\caption{
Computational workflow of the superconducting slave-boson calculation.
For each $(n,U,T)$ point, the paramagnetic SRIKR saddle point and the
Gaussian spin and charge fluctuations are recalculated. The dominant
static magnetic wave vector and the corresponding effective interaction
$U_{\mathrm{eff}}(n,U,T)$ are then determined. The resulting static or
dynamical pairing kernel is inserted into the anisotropic gap equation,
which is iterated until self-consistency is achieved. The entire
procedure is repeated during the temperature sweep until the
superconducting transition temperature is located.
}
\label{fig:computational_workflow}

\end{figure*}

\section{Frequency-dependent gap profile}
\label{sec:supp_freq_gap}

\begin{figure}[t]
    \centering
    \includegraphics[width=0.5\linewidth]{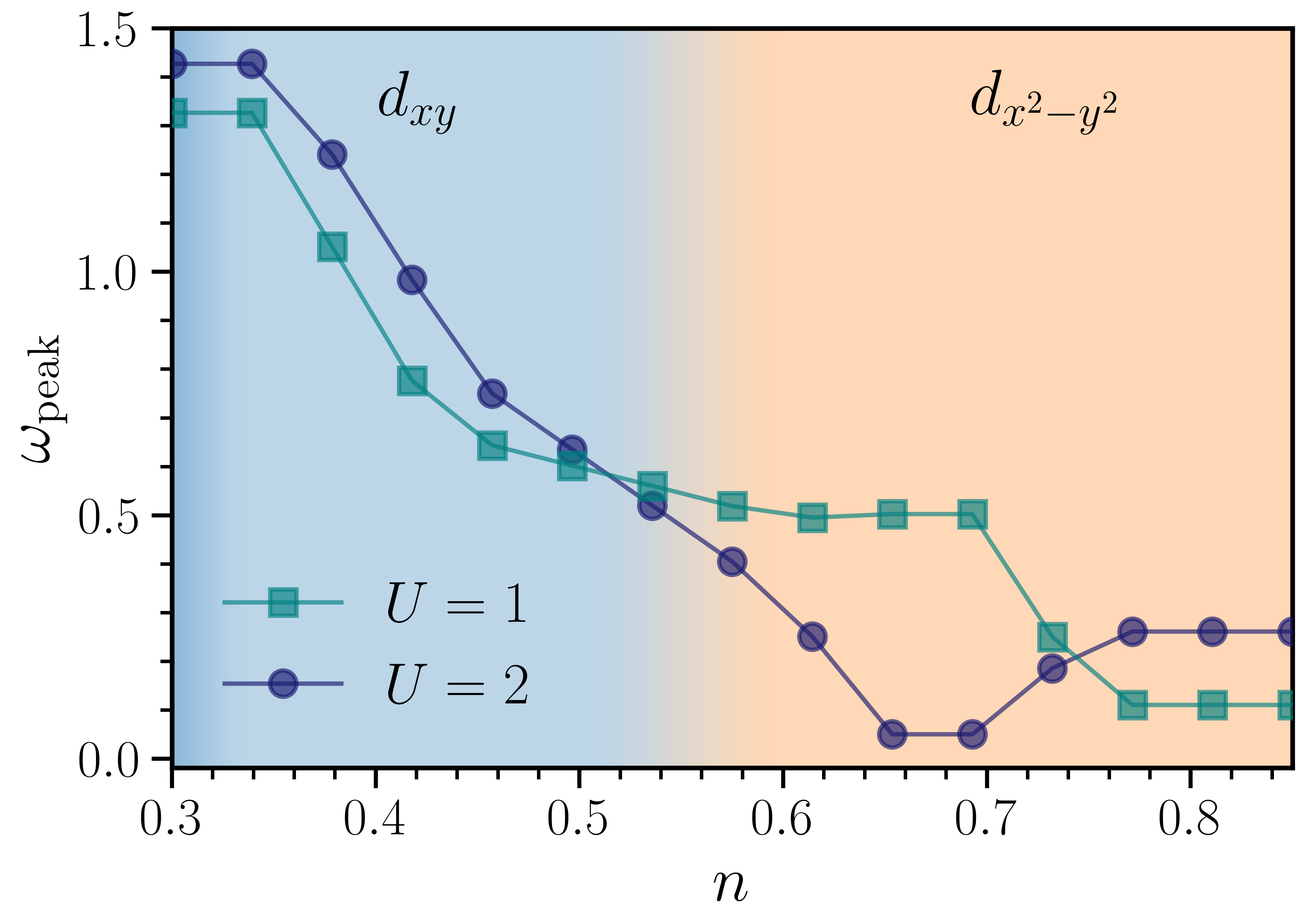}
    \caption{
    \textbf{FLEX+RPA result for the real-frequency gap peak scale.}
    Peak position $\omega_{\mathrm{peak}}$ of
    $\operatorname{Re}\Delta(\omega)$ obtained from the FLEX+RPA
    calculation for $t'=-0.2$ and $U=1,2$. The peak scale exhibits an
    overall softening with increasing filling and a corresponding change
    across the $d_{xy}\!\to d_{x^2-y^2}$ symmetry transition, consistent
    with the trend obtained within the slave-boson approach.
    }
    \label{fig:5}
\end{figure}

This section summarizes the Pad\'e analytic continuation used in the main text to extract real-frequency information
from the Matsubara-axis gap solutions.

\subsection{Pad\'e analytic continuation}

Starting from the Matsubara data
$\{z_n=i\omega_n,\Delta(i\omega_n)\}$ at fixed momentum $\mathbf{k}_0$, we approximate the gap by a Pad\'e
rational function
\begin{equation}
P_N(z)=\frac{A_N(z)}{B_N(z)},
\label{eq:supp_pade}
\end{equation}
constructed to interpolate the computed Matsubara points. The real-frequency gap is then obtained as
\begin{equation}
\Delta(\omega)=\lim_{\eta\to 0^+}P_N(\omega+i\eta),
\label{eq:supp_real_gap}
\end{equation}
with a small finite $\eta$ retained in practice for numerical stability. Following Refs.~\cite{Vidberg1977s,Baker1975e}, the Pad\'e approximant is represented in continued-fraction form.
Defining coefficients $\{a_n\}$ and recursion polynomials $A_n(z)$, $B_n(z)$, the stable continued-fraction
construction reads
\[
A_{n+1}(z)=A_n(z)+(z-z_n)a_{n+1}A_{n-1}(z),
\qquad
B_{n+1}(z)=B_n(z)+(z-z_n)a_{n+1}B_{n-1}(z).
\]
For numerical stability one propagates the ratios
\[
\bar A_n(z)=\frac{A_n(z)}{A_{n-1}(z)},
\qquad
\bar B_n(z)=\frac{B_n(z)}{B_{n-1}(z)},
\]
so that the Pad\'e approximant can be built recursively while controlling round-off errors at large order. To assess the stability of the analytically continued real-frequency
gap, we repeated the Padé continuation for several Padé orders \(N\),
broadening parameters \(\eta\), and real-frequency fitting windows.
The qualitative filling dependence of \(\omega_{\rm peak}\) remains
unchanged under moderate variations of these numerical parameters.
In particular, the downward shift of \(\omega_{\rm peak}\) with
increasing filling and the change in slope near the
\(d_{xy}\rightarrow d_{x^2-y^2}\) crossover are robust for both SB \& FLEX+RPA (Fig.~\ref{fig:5}).

\vspace{1em}
\begin{center}
\rule{0.6\textwidth}{0.4pt}
\end{center}



\suppTOCoff
\end{document}